\documentclass[sigconf,authorversion]{acmart}

\usepackage[colorinlistoftodos,prependcaption,disable]{todonotes} %put in disable to make them invisible
\presetkeys%
    {todonotes}%
    {inline,backgroundcolor=orange}{}

\usepackage{caption}
\usepackage{subcaption}

\usepackage{placeins} % -> use \FloatBarrier

\usepackage{stfloats}
\usepackage{float}

\usepackage{tabularx}
\usepackage{color, colortbl}

\usepackage{bytefield}

\usepackage{tikz}
\usetikzlibrary{shapes,arrows,calc,positioning,matrix}
\usetikzlibrary{decorations.pathreplacing}

\usepackage{adjustbox}
\usepackage{array}
\newcolumntype{R}[2]{%
    >{\adjustbox{angle=#1,lap=\width-(#2)}\bgroup}%
    l%
    <{\egroup}%
}
\usepackage{wasysym}

\usepackage[binary-units]{siunitx}
\definecolor{darkred}{RGB}{212,0,16}
\definecolor{darkgreen}{RGB}{46,180,38}
\definecolor{darkergreen}{RGB}{6,98,1}
\definecolor{darkblue}{RGB}{0,73,218}
\definecolor{sorange}{rgb}{0.95, 0.57, 0}
\colorlet{orange}{sorange}

\usepackage{listings}
\lstdefinelanguage{ASM}{
    morekeywords={b, ble, blt, bne, bx, bl, ldr, str, push, pop, mov, add, sub, beq, cmp},
    keywordstyle=\color{darkblue},
    sensitive=false, % keywords are not case-sensitive
    morecomment=[l]{//}, % l is for line comment
    morecomment=[l]{;}, % l is for line comment
    morecomment=[s]{/*}{*/}, % s is for start and end delimiter
    commentstyle=\color{darkgreen},
    morestring=[b]", % defines that strings are enclosed in double quotes
} %
\lstdefinelanguage{none}{
  identifierstyle=
}

\usepackage{xspace}

\usepackage[nolist]{acronym}

\acrodefplural{PoC}[PoCs]{Proofs of Concept}

\begin{acronym}
\acro{SDR}{Software-Defined Radio}
\acro{AGC}{Automatic Gain Control}
\acro{GCI}{Global Coexistence Interface}
\acro{ECI}{Enhanced Coexistence Interface}
\acro{AFH}{Adaptive Frequency Hopping}
\acro{FEM}{Front-End Module}
\acro{SP3T}{Single Pole, Triple Throw}
\acro{HCI}{Host Controller Interface}
\acro{A2DP}{Advanced Audio Distribution Profile}
\acro{SCO}{Synchronous Connection-Oriented}
\acro{GPIO}{General Purpose Input Output}
\acro{ASLR}{Address Space Layout Randomization}
\acro{LCP}{Link Control Protocol}
\acro{LNA}{Low-Noise Amplifier}
\acro{EIR}{Extended Inquiry Response}
\acro{CRC}{Cyclic Redundancy Check}
\acro{RTOS}{Real-Time Operating System}
\acro{QEMU}{Quick Emulator}
\acro{UART}{Universal Asynchronous Receiver Transmitter}
\acro{MitM}{Machine-in-the-Middle}
\acro{MAC}{Media Access Control}
\acro{BLE}{Bluetooth Low Energy}
\acro{ACL}{Asynchronous Connection-Less}
\acro{DSP}{Digital Signal Processing}
\acro{SDIO}{Secure Digital Input Output}

\acro{BCS}{Bluetooth Core Scheduler}
\acro{ELF}{Executable and Linking Format}
\acro{GIAC}{Global Inquiry Access Code}
\acro{JSON}{JavaScript Object Notation}
\acro{L2CAP}{Logical Link Control and Adaptation Protocol}
\acro{PTM}{Pseudo Terminal Master}
\acro{PTS}{Pseudo Terminal Slave}
\acro{QEMU}{Quick Emulator}
\acro{RCE}{Remote Code Execution}
\acro{RFU}{Reserved for Future Use}
\acro{SVC}{Supervisor Call}
\acro{UART}{Universal Asynchronous Receiver Transmitter}
\acro{WICED}{Wireless Internet Connectivity for Embedded Devices}
\acro{MMIO}{Memory Mapped Input/Output}
\acro{LE}{Bluetooth Low Energy}
\acro{IoT}{Internet of Things}
\acro{IDE}{Integrated Development Environment}
\acro{ARM}{Advanced RISC Machine}
\acro{LM}{Link Manager}

\acro{RTOS}{Real-Time Operating System}
\acro{DMA}{Direct Memory Access}
\acro{RXDMA}{Receive Direct Memory Access}
\acro{NVRAM}{Non-Volatile Random-Access Memory}
\acro{ROM}{Read-Only Memory}
\acro{Rx}{Receive}
\acro{Tx}{Transmit}
\acro{SPI}{Serial Peripheral Interface}
\acro{MSP}{Main Stack Pointer}
\acro{PSP}{Process Stack Pointer}
\acro{RF}{Radio Frequency}
\acro{BLOB}{Binary Large Object}
\acro{SCO}{Synchronous Connection Oriented}
\acro{UAF}{Use-After-Free}
\acro{FHS}{Frequency Hop Sync}
\acro{CRC}{Cyclic Redundancy Check}
\acro{PDU}{Protocol Data Unit}
\acro{NFC}{Near Field Communication}
\acro{JPEG}{Joint Photographic Experts Group}
\acro{LPE}{Local Privilege Escalation}
\acro{DEP}{Data Execution Prevention}
\acro{XN}{eXecute Never}
\acro{LCP}{Link Control Protocol}
\acro{GATT}{Generic Attribute}
\acro{PoC}{Proof of Concept}
\acro{EDR}{Enhanced Data Rate}
\acro{MWS}{Mobile Wireless Standards}
\acro{HAL}{Hardware Abstraction Layer}
\acro{LR}{Link Register}
\acro{eSIM}{embedded-SIM}
\acro{RSP}{Remote SIM Provisioning}
\acro{PC}{Program Counter}
\acro{MD}{More Data}
\acro{LLID}{Logical Link Identifier}
\acro{SFI}{Serial Flash Interface}
\acro{EEPROM}{Electrically Erasable Programmable Read-Only Memory}
\acro{TOFU}{Trust On First Use}
\acro{CVE}{Common Vulnerabilities and Exposure}
\acro{PCIe}{Peripheral Component Interconnect Express}
\acro{SECI}{Serial Enhanced Coexistence Interface}
\acro{HID}{Human Interface Device}
\acro{DoS}{Denial of Service}
\acro{PTA}{Packet Traffic Arbitration}
\acro{AWMA}{Alternating Wireless Medium Access}
\acro{ECDH}{Elliptic-Curve Diffie--Hellman}
\acro{BT}{Bluetooth Classic}
\acro{SSP}{Secure Simple Pairing}
\acro{SC}{Secure Connections}
\acro{MIC}{Message Integrity Check}

\acro{SE}{Secure Element}
\acro{SEP}{Secure Enclave Processor}
\acro{UWB}{Ultra-wideband}
\acro{DCK}{Digital Car Key}
\acro{LPM}{Low-Power Mode}
\acro{PMU}{Power Management Unit}
\acro{FPB}{Flash Patch and Breakpoint Unit}
\acro{I2C}[I$^2$C]{Inter-Integrated Circuit}
\acro{XPC}{Cross-Process Communication}
\acro{TLV}{Type Length Value}
\acro{AP}{Application Processor}
\acro{AoP}{Always-on Processor}
\acro{DYLD}{Dynamically Loaded Library}
\acro{DSP}{Digital Signal Processor}
\acro{OTA}{Over-the-Air}
\acro{NSA}{National Security Agency}

\end{acronym}

\usepackage{mdframed}

\AtBeginDocument{%
  \providecommand\BibTeX{{%
    \normalfont B\kern-0.5em{\scshape i\kern-0.25em b}\kern-0.8em\TeX}}}

\copyrightyear{2022}
\acmYear{2022}
\setcopyright{acmlicensed}
\acmConference[WiSec '22]{Proceedings of the 15th ACM Conference on Security and Privacy in Wireless and Mobile Networks}{May 16--19, 2022}{San Antonio, TX, USA}
\acmBooktitle{Proceedings of the 15th ACM Conference on Security and Privacy in Wireless and Mobile Networks (WiSec '22), May 16--19, 2022, San Antonio, TX, USA}
\acmPrice{15.00}
\acmDOI{10.1145/3507657.3528547}
\acmISBN{978-1-4503-9216-7/22/05}

\begin{document}

\title[Evil Never Sleeps]{Evil Never Sleeps: \\ When Wireless Malware Stays On After Turning Off iPhones}


\author{Jiska Classen}
\affiliation{%
  \institution{Secure Mobile Networking Lab, TU Darmstadt}
%  \institution{TU Darmstadt}
%  \streetaddress{Pankratiusstraße 2}
%  \city{Darmstadt}
  \country{Germany}
%  \postcode{64293}
}
\email{jclassen@seemoo.de}

\author{Alexander Heinrich}
\affiliation{%
  \institution{Secure Mobile Networking Lab, TU Darmstadt}
%  \institution{TU Darmstadt}
%  \streetaddress{Pankratiusstraße 2}
%  \city{Darmstadt}
  \country{Germany}
%  \postcode{64293}
}
\email{aheinrich@seemoo.de}

\author{Robert Reith}
\affiliation{%
  \institution{Secure Mobile Networking Lab, TU Darmstadt}
%  \institution{TU Darmstadt}
%  \streetaddress{Pankratiusstraße 2}
%  \city{Darmstadt}
  \country{Germany}
%  \postcode{64293}
}
\email{rreith@seemoo.de}

\author{Matthias Hollick}
\affiliation{%
  \institution{Secure Mobile Networking Lab, TU Darmstadt}
%  \institution{TU Darmstadt}
%  \streetaddress{Pankratiusstraße 2}
%  \city{Darmstadt}
  \country{Germany}
%  \postcode{64293}
}
\email{mhollick@seemoo.de}

%!TEX root = ../lpm.tex

\begin{abstract}

When an iPhone is turned off, most wireless chips stay on.
For instance, upon user-initiated shutdown, the iPhone remains locatable via the
Find My network. If the battery runs low, the iPhone shuts down
automatically and enters a power reserve mode. Yet, users can still access
credit cards, student passes, and other items in their Wallet.
We analyze how Apple implements these standalone wireless features, working while iOS is not running, and determine
their security boundaries.
On recent iPhones, Bluetooth, \ac{NFC}, and \ac{UWB} keep running after power off,
and all three wireless chips have direct access to the secure element.
As a practical example what this means to security, we demonstrate
the possibility to load malware onto a Bluetooth chip that is executed while
the iPhone is off.

\end{abstract}





\maketitle

%!TEX root = ../lpm.tex

\section{Introduction}

\begin{figure}[!b]
\vspace{-1em} %TODO
\includegraphics[width=1.0\columnwidth]{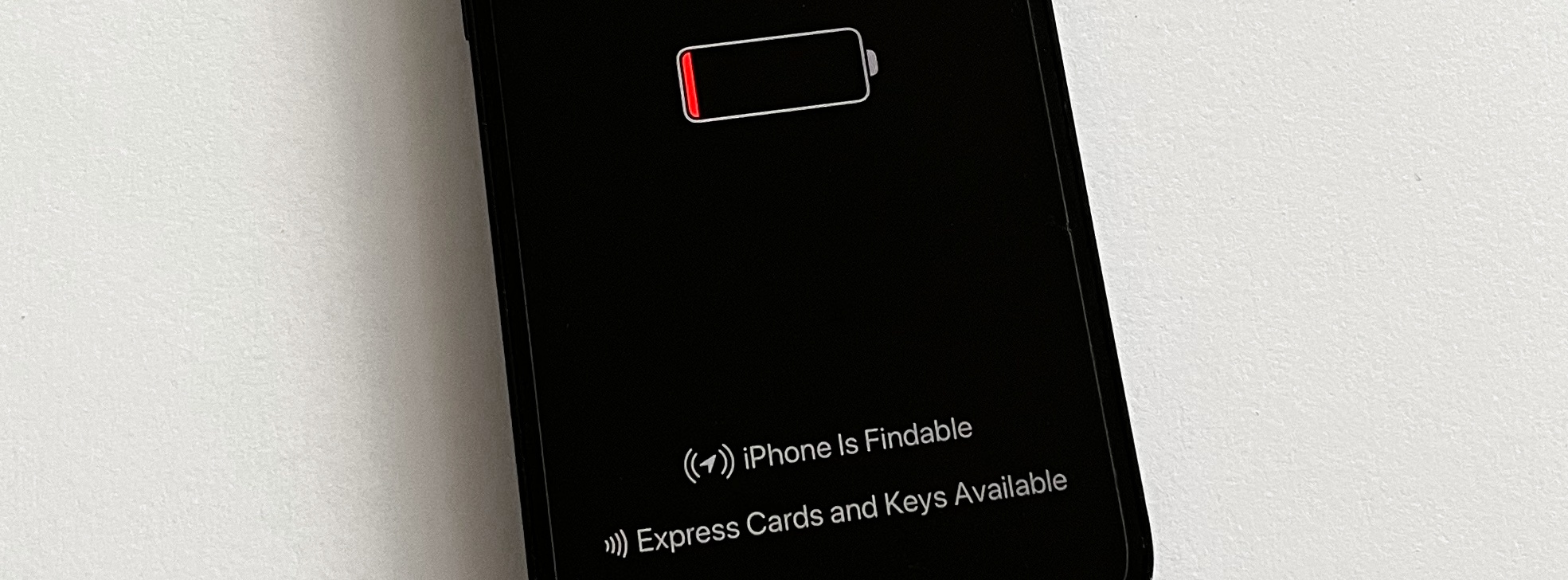}
\caption{Find My and Express Cards in power reserve mode.}
\label{fig:express_pwr}
\end{figure}

Wireless chips on the iPhone can run in a so-called \ac{LPM}.
Note that \ac{LPM} described in this paper is different from the energy saving 
mode indicated by a yellow battery icon. In \ac{LPM},
the iPhone does not react to tapping the screen or shaking.
This mode is either activated when the user switches off their
phone or when iOS shuts down automatically due to low battery.
If the iPhone was switched off by the user, it turns on when pressing the power button.
In battery-low power reserve mode, pressing the power button only activates the screen for a few seconds.
It indicates the low battery and lists the currently active \ac{LPM} features, as shown in \autoref{fig:express_pwr}.

\ac{LPM} features increase the user's security, safety, and convenience in most situations.
Find My is available after user-initiated and battery-low shutdown. A user who loses their
iPhone while it is turned off can locate it using the Bluetooth-based Find My network~\cite{heinrich2021findmy}.
Express Mode supports selected student, travel, and credit cards, as well as digital keys
to be used faster without additional authentication by the user.
Upon battery-low shutdown, these cards and keys can be used for up to \SI{5}{\hour}~\cite{expressmode}.
Initially, this only required the \ac{NFC} to stay on.
Instead, the new \ac{DCK} 3.0 protocol uses Bluetooth and \ac{UWB},
and it also supports Express Mode in power reserve~\cite{appleuwbcarwwdc}.
Express Mode protects users against locking themselves out of their cars and homes and being unable to
make payments.

\ac{LPM} support is implemented in hardware. The \ac{PMU} can turn on chips individually. The Bluetooth
and \ac{UWB} chips are hardwired to the \ac{SE} in the \ac{NFC} chip, storing secrets that should be available in \ac{LPM}.
Since \ac{LPM} support is implemented in hardware, it cannot be removed by changing
software components. 
As a result, on modern iPhones, wireless chips can no longer be trusted to be turned off after shutdown.
This poses a new threat model.
Previous work only considered that journalists are not safe against espionage when
enabling airplane mode in case their smartphones were compromised~\cite{Huang2016Against}.
\ac{LPM} is significantly more stealthy than a fake power off that only
disables the screen. Even though the \ac{NSA} used fake power off on
smart TVs for espionage~\cite{wikileaks}, high battery drainage would
be noticeable and could reveal such implants on mobile devices.
We show that \ac{LPM} is a relevant attack surface that has to be considered
by high-value targets such as journalists, or that can be weaponized to build wireless
malware operating on shutdown iPhones.

We are the first to analyze Apple's \ac{LPM} features. We reverse-engineer multiple
wireless daemons and firmware components to systematically analyze \ac{LPM}.
Our key contributions are:
\begin{itemize}
\item We perform a security analysis of new \ac{LPM} features introduced in iOS 15,
\item we identify flaws in the Find My \ac{LPM} implementation, limiting the total advertisement time to \SI{24}{\hour} even after reboot,
\item we design, implement, and open-source tooling to analyze and modify Bluetooth firmware on recent iPhones, and
\item we demonstrate that Bluetooth \ac{LPM} firmware can be modified to run malware.
\end{itemize}

We responsibly disclosed all issues to Apple. They read the paper prior to publication but had no feedback on the paper's contents.
The tooling required for firmware analysis and modification on modern iPhones is available as part of the \emph{InternalBlue}~\footnote{https://github.com/seemoo-lab/internalblue} and
\emph{Frankenstein}~\footnote{https://github.com/seemoo-lab/frankenstein} repositories.

The remainder of this paper is structured as follows.
Apple describes how \ac{NFC} \ac{LPM} is supposed to work, which we summarize in \autoref{sec:known_lpm}.
Then, we explain undocumented \ac{LPM} Bluetooth and \ac{UWB} features in \autoref{sec:new_lpm}.
Based on this knowledge, we analyze threats posed by \ac{LPM} in \autoref{sec:threat}.
In \autoref{sec:malware}, we demonstrate that it is possible to modify the Bluetooth firmware loaded while the chip is in \ac{LPM}.
Our reverse-engineering methods are described in \autoref{sec:reversing} to ensure reproducibility of our results.
We discuss related work in \autoref{sec:related} and conclude our paper in \autoref{sec:conclusion}.

%!TEX root = ../lpm.tex

\section{NFC Low-Power Mode}
\label{sec:known_lpm}

From a hardware perspective, enabling Bluetooth, \ac{UWB}, or other chips like \ac{NFC} while the remaining iPhone
is off are very similar.
Bluetooth and \ac{UWB} \ac{LPM} features are undocumented and have to the best of our knowledge not been investigated, yet.
However, Apple's platform security guide~\cite{appleplatformsec} documents an \ac{NFC} \ac{LPM} feature called
Express Card, introduced in iOS 12.
Hardware components supporting Express Cards and enforcing security boundaries are shown in \autoref{fig:se_sep}.
We describe these in the following.

\tikzset{>=latex}
\begin{figure}[!t]
%\vspace{-0.5em} %TODO
	\center
	\begin{tikzpicture}[minimum height=0.55cm, scale=0.8, every node/.style={scale=0.8}, node distance=0.7cm]

    %NFC+SE chip
	\filldraw[fill=gray!20,draw=gray,thick, align=center](0,0) rectangle node (nfc) {} ++(4.25,2.75);
	\node[anchor=north,align=center, above=of nfc.north,yshift=0.5em] (chiptxtb) {\textbf{NFC with Secure Element}}; 
	%\filldraw[fill=darkgreen!10,draw=darkgreen,thick, align=left](1,0.25) rectangle node (se) {\includegraphics[width=0.7em]{pics/contactless_symbol.pdf} Apple Pay \\ \includegraphics[width=0.7em]{pics/car_symbol.pdf} Car Keys \\~\\ \includegraphics[width=0.7em]{pics/express_symbol.pdf} Express Cards} ++(3,1.75);
	%\draw[-, darkgreen] (1, 1) -- (4, 1);  % separate SE into two parts
	\fill[fill=darkgreen!10](1,0.25) rectangle node (se) {} ++(3,1.75);
	\node[anchor=north,align=center, above=of se.north,yshift=-1em,color=darkgreen] (chiptxtb) {Secure Element}; 
	%separate SE overlays
	\node[align=left, anchor=west] at (1.3, 0.6) {\includegraphics[width=0.7em]{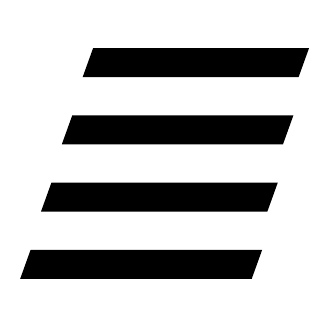} Express Cards};
	\fill[fill=darkgreen!25, align=left](1,0.9) rectangle node (applet) {} ++(3,1.1);
	\draw[-, darkgreen] (1, 0.9) -- (4, 0.9);  % separate SE into two parts
	\node[align=left, anchor=west] at (1.3, 1.4) {\includegraphics[width=0.7em]{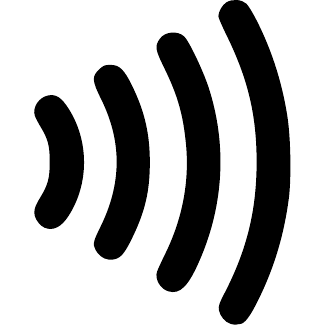} Apple Pay \\ \includegraphics[width=0.7em]{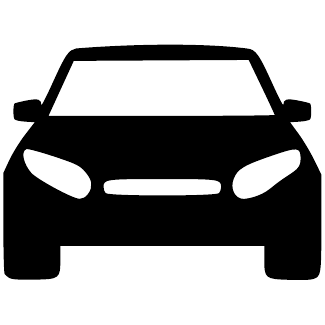} Car Keys};
	\draw[draw=darkgreen,thick](1,0.25) rectangle node (seoutline) {} ++(3,1.75);

	%NFC antenna
	\begin{scope}[shift={(0,1)}]
    \draw[-, gray] (0,-0.25) -- (-0.75, -0.25) -- (-0.75, 1.2) -- (-0.55, 1.5) -- (-0.95, 1.5) -- (-0.75, 1.2);
    \draw[-, gray, dotted] (0, -0.25) -- (1, -0.25); % continue line into secure element
    \node[align=center,darkgray] at (0.5, -0.5) {LPM};
    \end{scope}

	\filldraw[fill=gray!20,draw=gray,thick, align=center](5.5,0) rectangle node (sep) {Authorization \\ (Touch ID / Face ID / PIN)} ++(4,2.75);
	\node[anchor=north,align=center, above=of sep.north,yshift=0em] (chiptxtb) {\textbf{Secure Enclave Processor}}; 
	\node[align=center] at (5.75, 1.5) (finger) {\includegraphics[width=0.9em]{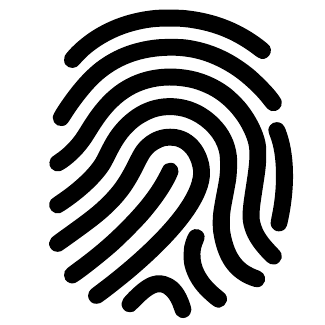}};
	
	% connection between se and sep
	
	\path[draw, thick] (4,1.5) edge node[xshift=0.1cm, yshift=0.75em] {Paired} (5.5,1.5);
    
	\end{tikzpicture}

%\vspace{-0.5em} %TODO
\caption{\acf{SE} and \acf{SEP} usage as documented by Apple.}
\label{fig:se_sep}
%\vspace{-1em} %TODO
\end{figure}

\subsection{Security Boundaries}

iOS implements secure wireless payments on an \ac{NFC} chip with a \acf{SE}.
The \ac{SE} is certified based on Common Criteria and runs a JavaCard platform~\cite[p. 140]{appleplatformsec}.
This platform executes applets, e.g., Apple Pay~\cite[p. 141]{appleplatformsec}, \ac{DCK}~\cite[p. 169]{appleplatformsec}, or other third-party applets.
During setup, applets are personalized. Personalization can have additional security boundaries, e.g., no credit card information is stored on a payment applet
but a revokable identifier known to the payment network~\cite[p. 145]{appleplatformsec}.
Applets, including their secrets, are separated from each other. They are stored within the \ac{SE} and do not leave it.
During a payment process or similar wireless transactions, the \ac{SE} within the \ac{NFC} chip replies directly
without forwarding this data to iOS~\cite[p. 141]{appleplatformsec}.
Thus, even if iOS or apps are compromised, credit cards and other keys cannot be stolen from the \ac{SE}~\cite{serelay}.

The \ac{SE} is connected to the \acf{SEP}. The \ac{SEP} authorizes payments by enforcing that the user authenticates
using Touch ID, Face ID, or a passcode~\cite[p. 141]{appleplatformsec}.
Note that the \ac{SEP} does not store data within the \ac{SE} but uses a separate secure storage component~\cite[p. 15]{appleplatformsec}. %TODO not sure if relevant, remove if space is needed
The \ac{SE} and \ac{SEP} are paired in-factory and, thus, can communicate encrypted and authenticated~\cite[p. 144]{appleplatformsec}.

\subsection{Express Cards and Power Reserve}

The Wallet app allows configuring selected \ac{NFC} credit, travel, and student cards as well as keys for Express Mode.
An Express Card does no longer require authorization via the \ac{SEP}~\cite[p. 153f]{appleplatformsec},
enabling fast and convenient payment without unlocking the iPhone.
Possessing the iPhone is the only authorization in Express Mode.
Skipping authorization means that neither the \ac{SEP} nor iOS are required to complete an \ac{NFC} transaction, and \ac{NFC} can operate standalone using \ac{SE} applets.

Once an iPhone runs out of battery and if the user has an Express Card, the iPhone shuts down but keeps the \ac{NFC} chip powered
for up to \SI{5}{\hour}~\cite{expressmode}. This enables the user to use their transit and credit cards
as well as car keys until they can charge their iPhone.
During this period, the display indicates Express Cards are enabled when pressing the power button, as shown in \autoref{fig:express_pwr}.
Express Cards are only enabled in power reserve mode.
A user-initiated shutdown also powers off \ac{NFC}~\cite[p. 28]{appleplatformsec}.

%!TEX root = ../lpm.tex

\section{Bluetooth and UWB LPM}
\label{sec:new_lpm}

iOS 15 introduces two new \ac{LPM} features:
\emph{(i)} Find My, Apple's \ac{BLE}-based offline finding network, and \emph{(ii)} \acf{DCK} 3.0 support,
which uses \ac{UWB} for a secure distance measurement. Thus, also the Bluetooth and the \ac{UWB}
chip are able to operate standalone while iOS is powered off. These capabilities are undocumented and have not been
researched before. To improve readability, we explain our reverse engineering process later in \autoref{sec:reversing},
and only describe how these modes work in the following.

\subsection{Find My Bluetooth Module}

\subsubsection{Working Principle}
Find My is an offline finding network, which has been reverse-engineered in detail~\cite{heinrich2021findmy}.
It locates iPhones, AirTags, and other Apple devices while they are not connected to the Internet.
To this end, offline devices in lost mode regularly send \ac{BLE} advertisements.
Devices with a network connection scan for these advertisements and report 
lost devices' locations to the legitimate owners.

The Find My protocol is end-to-end encrypted, anonymous, and privacy preserving.
This security guarantee is based on a device-specific master beacon key.
The master beacon key is synchronized with the user's iCloud keychain, allowing
access as long as the user can log in to iCloud. The keychain is also stored locally,
and access to its secret key requires a roundtrip through the \ac{SEP}~\cite[p. 96]{appleplatformsec}. Based on the master beacon key, a rolling
sequence of private/public key pairs is generated. This sequence is fixed forever, 
and each key is only valid during its time slot.
Linking public keys in this
sequence to each other requires knowledge of the master beacon key. The device
broadcasts the current public key as \ac{BLE} advertisement, and a portion of the
key also sets the MAC address to a random value.
Any Internet-connected device that observes a Find My \ac{BLE}
advertisement can encrypt its current approximate location with the public key and report it to Apple.
Only the legitimate owner of the offline device possesses the matching private key to decrypt the location report.
The location report does not include the reporter's identity.
Public keys roll every \SI{15}{\minute} on iPhones, which limits smartphone
tracking via \ac{BLE} advertisements.

AirTags work similar to Find My on iPhones. When they lose the Bluetooth connection to the owner's iPhone,
they start sending advertisements.
On AirTags, the advertisement public key only rolls every \SI{24}{\hour}.
This way, it can be detected when an AirTag follows the same person, and an
anti-stalking warning can be displayed on the tracked person's smartphone~\cite{airguard}. %TODO in final version, link alex' final paper
However, a longer rolling
time limits the legitimate user's privacy.
A \SI{24}{\hour} time interval might be a technical requirement on
AirTags, which do not have a precise time source. Drifting away
from a shorter \SI{15}{\min} window would mean that they could no longer be found
when querying Apple's servers.

\begin{figure}[!b]
%  trim={<left> <lower> <right> <upper>}
%\includegraphics[trim={0 65cm 0 0},clip, width=0.7\columnwidth]{pics/poweroff.png}
\includegraphics[width=0.6\columnwidth]{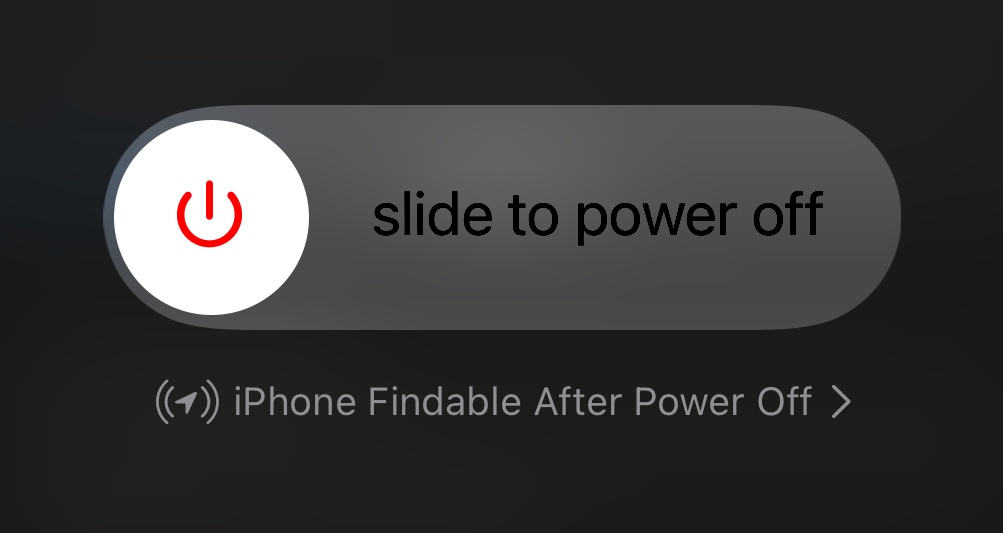}

\caption{Find My shutdown dialogue on iOS 15.}
\label{fig:findmy}
\end{figure}

\subsubsection{\acl{LPM} Setup}

When Find My is supported after power off, this is shown as part of the shutdown dialogue.
Users can change this setting during each manual power off procedure as depicted in \autoref{fig:findmy}. If the iPhone runs out of battery and enters
power reserve mode, Find My is enabled automatically, similar to Express Cards in the \ac{NFC} chip.

\ac{LPM} support requires a Bluetooth firmware that can send \ac{BLE} advertisements while iOS is off.
Broadcom Bluetooth chips can either be configured to interact with a host, such as iOS, or
to run as standalone application, such as IoT devices. When entering \ac{LPM}, the iOS Bluetooth stack is terminated and
the Bluetooth chip is reset.
Then, multiple \ac{HCI} commands configure Find My parameters. The last \ac{HCI} command disables
\ac{HCI}, thereby stopping all communication with iOS. The firmware starts a standalone Find My advertisement thread.
The \ac{PMU} is advised to keep the Bluetooth chip on despite shutting iOS down.

The Find My configuration \ac{HCI} commands are very flexible. They allow setting multiple public keys for a short and a long rotation
interval. The rotation interval duration and the total number of keys are variable parameters. As of iOS 15.3, \num{96}
short interval (\SI{15}{\min}) and \num{0} long interval (\SI{24}{\hour}) keys are set. Thus, the standalone Bluetooth app
can only send Find My advertisements for up to \SI{24}{\hour}. In power reserve mode, entered due to low-battery shutdown,
Find My stops even earlier, after slightly more than \SI{5}{\hour}. This property is opaque to users, who might expect that their iPhone
remains findable for a longer period of time.

Only configuring a limited amount of short interval public keys has multiple security advantages.
Advertisement linkability is reduced by the short rotation time, which increases privacy~\cite{mayberry}.
This also means that the iPhone will not trigger anti-stalking features, thereby hiding a
potentially lost iPhone from thieves. Most importantly, the master beacon key stays within the keychain
and is not shared with the \ac{SE}, Bluetooth chip, or Bluetooth daemon.

\subsection{Digital Car Key 3.0 Bluetooth and UWB Modules}
\label{ssec:dck3}

\subsubsection{UWB as NFC Successor for Car Keys}

In addition to \ac{NFC} cards, iOS also supports car keys.
Since they are digital, they can have other properties than regular car keys.
For example, they can be individualized per user, including speed limits for young drivers,
and shared between iCloud users~\cite{bmwdigitalkey,appleuwbcarwwdc}.
While having \ac{DCK} support on iOS is convenient for users, this
also adds new attack vectors to car theft. On the implementation side,
the main risk is on car manufacturers---a compromised \ac{DCK} or failures in its implementation can be used to steal
a car but not to steal an iPhone. Unlike digital payments, physical car theft cannot be undone.
This risk motivated car manufacturers to move from \ac{DCK} 2.0, which is based on
\ac{NFC}, to \ac{DCK} 3.0.

Even though the acronym \acf{NFC} suggests that only two nearby communication parties
can perform transactions successfully, \ac{NFC} and older car key technologies are prone to relay attacks~\cite{nxpdigitalkey,pkesshortening}.
In a relay attack, signals are forwarded over a larger distance than intended.
Relay attacks pose a significant threat to applications like \acp{DCK},
because they allow car access and theft while the user is not nearby.
Short distances can be reduced by amplifying an existing signal. Long distances require
decoding a signal into packets and forwarding these.
The first practical low-cost tool for performing packet-based \ac{NFC} relay attacks has been
published in 2011~\cite{relay2011}, and similar tools supporting up-to-date \ac{NFC}
technologies are still maintained~\cite{proxmark, nfcgatepaper}.

While the \ac{DCK} 2.0 specification relies on \ac{NFC}~\cite{digitalkeywhitepaper},
\ac{DCK} 3.0 adds support for secure ranging based on \ac{UWB}~\cite{digitalkey30press}.
Even though \ac{UWB} was marketed to support secure ranging, the \ac{UWB} distance
measuring fundamentals are flawed~\cite{wisecuwb}, and low-cost distance shortening
attacks against Apple's \ac{UWB} chip of up to \SI{12}{\meter} were demonstrated in practice~\cite{ghostpeak}.
We assume that these flaws were unknown during the \ac{DCK} 3.0 design phase, and \ac{UWB}
still provides better ranging security than \ac{NFC}.

% https://developer.apple.com/videos/play/wwdc2021/10084/
% at 6:11 - yes that's also available with LPM!

\tikzset{>=latex}
\begin{figure}[!t]
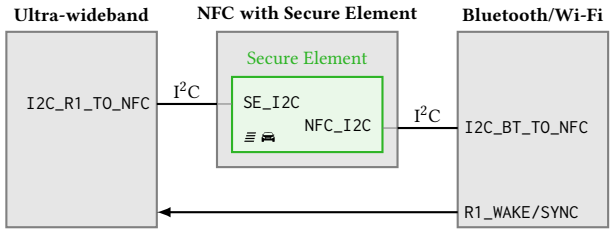

%\vspace{-0.5em} %TODO
	\center
	\begin{tikzpicture}[minimum height=0.55cm, scale=0.8, every node/.style={scale=0.8}, node distance=0.7cm]

    %NFC+SE chip
	\filldraw[fill=gray!20,draw=gray,thick, align=center](0,0.5) rectangle node (nfc) {} ++(3,2.75-0.5);
	\node[anchor=north,align=center, above=of nfc.north,yshift=0em] (chiptxtb) {\textbf{NFC with Secure Element}}; 
	\filldraw[fill=darkgreen!10,draw=darkgreen,thick, align=left](0.25,0.75) rectangle node[] (se) {\texttt{SE\_I2C} \\ \hspace*{3em} \texttt{NFC\_I2C}} ++(2.5,1.75-0.5);
	\node[] at (0.7, 1) {\includegraphics[width=0.7em]{pics/express_symbol.pdf} \includegraphics[width=0.7em]{pics/car_symbol.pdf}};
	\node[anchor=north,align=center, above=of se.north,yshift=-2em,color=darkgreen] (chiptxtb) {Secure Element};

	\filldraw[fill=gray!20,draw=gray,thick, align=left](4,-0.5) rectangle node (bt) {} ++(2.5,3.25); %TIME_SYNC and WAKE
	\node[align=left, anchor=west] at (4, -0.25) {\texttt{R1\_WAKE/SYNC}};
	\node[align=left, anchor=west] at (4, 1.15) {\texttt{I2C\_BT\_TO\_NFC}};
	\node[anchor=north,align=center, above=of bt.north,yshift=1.5em] (chiptxtb) {\textbf{Bluetooth/Wi-Fi}};

	\filldraw[fill=gray!20,draw=gray,thick, align=right](-3.5,-0.5) rectangle node (uwb) {} ++(2.5,3.25);
	\node[align=right, anchor=east] at (-1, 1.55) {\texttt{I2C\_R1\_TO\_NFC}};
	\node[anchor=north,align=center, above=of uwb.north,yshift=1.5em] (chiptxtb) {\textbf{\acl{UWB}}};

	% connection between se and bt/r1
	\begin{scope}[shift={(0,0.25)}]
	\path[draw, thick] (3,0.9) edge node[yshift=0.75em] {I$^2$C} (4,0.9);
	\path[draw, gray] (2.75,0.9) -- (3,0.9);
	\path[draw, thick] (-1,1.3) edge node[yshift=0.75em] {I$^2$C} (0,1.3);
	\path[draw, gray] (0,1.3) -- (0.25,1.3);
	\end{scope}
	
	% wake arrow
	
	\path[->, draw, thick] (4, -0.25) --  (-1, -0.25);
    
	\end{tikzpicture}

%\vspace{-0.5em} %TODO
\caption{New \acf{SE} and inter-chip interfaces present in the iPhone 11, 12, and 13 series.}
\label{fig:se_bt_uwb}
%\vspace{-1em} %TODO
\end{figure}

\subsubsection{Working Principle}

The \ac{DCK} 3.0 protocol uses both, \ac{BLE} and \ac{UWB}.
\ac{BLE} sets up the initial connection and authentication.
Then, \ac{UWB} is used for fine ranging secured with a symmetric key, but without data transfer.
Thus, \ac{BLE} and \ac{UWB} both require access to secret key material.
The \ac{DCK} 3.0 specification describing precise steps of the \ac{BLE} and \ac{UWB} protocols
is exclusively available to members of the Car Connectivity Consortium.
Marketing materials only contain rudimentary information~\cite{digitalkeywhitepaper, digitalkey30press, nxpdigitalkey, bmwdigitalkey}.
The most detailed information was released in a presentation by Apple, which
confirms that \ac{DCK} 3.0 is supported in power reserve mode~\cite{appleuwbcarwwdc}.

Apple prepared hardware components for \ac{DCK} 3.0 since the iPhone 11.
On iPhones with \ac{DCK} 3.0 support, the \ac{SE} is hardwired to the Bluetooth and \ac{UWB} chips, as shown
in \autoref{fig:se_bt_uwb}. Thus, similar to \ac{NFC} transactions using the \ac{SE},
replies generated by the \ac{SE} are sent directly over-the-air without being
shared to iOS. This ensures the same security boundaries as \ac{DCK} 2.0,
while adding support for secure ranging.

\subsubsection{\acl{LPM} Setup}

When the iPhone shuts down due to low battery, and a car key is configured,
\ac{UWB} Express Mode is enabled. This mode is very similar to the \ac{NFC} Express Mode.

Initial \ac{DCK} 3.0 protocol steps only require \ac{BLE}.
Hence, \ac{UWB} can go to sleep, but a \ac{BLE} application keeps running.
The Bluetooth stack is terminated and a standalone
application is started on the Bluetooth chip, similar to the Find My \ac{LPM}.
Both applications, \ac{DCK} and Find My, run in the same thread.
Applications are only initialized and executed when certain flags are set,
meaning that the same firmware image can be used even if only one application is active.
The \ac{DCK} application scans for other \ac{BLE} devices, runs a \ac{GATT}
service for data exchange over \ac{BLE}, and retrieves key material from the \ac{SE}
using the \ac{I2C} protocol.

In contrast to the Bluetooth chip, the \ac{UWB} chip is sleeping.
The daemon responsible for \ac{UWB} sends
a special \ac{LPM} enabling and configuration packet to the chip.
Once secure ranging is needed, the Bluetooth chip sends a wake signal
over the hardwired connection shown in \autoref{fig:se_bt_uwb}.
Timing between Bluetooth and \ac{UWB} is synchronized to reduce
active receiver times to the expected incoming packet window~\cite{appleuwbcarwwdc}. % at 13:05
The combination of both saves a lot of power on the \ac{UWB} chip, since it is
sleeping most of the time, and even when active only listens within very limited time windows.

\subsection{Supported Devices}

Bluetooth and \ac{UWB} \ac{LPM} support are mainly driven by \ac{DCK} 3.0 integration.
Only iPhones with \ac{DCK} 3.0 support also have Bluetooth and \ac{UWB} \ac{LPM}, as listed in \autoref{tab:chip_list}.
The Bluetooth chip in the iPhone SE 2020 is the same as in the iPhone 11.
Find My could run standalone on this chip while powered by the \ac{PMU}.
However, as of iOS 15.2, the firmware on the iPhone SE 2020 is missing the \ac{LPM} thread.
It is unclear if this feature is missing intentionally or if a firmware supporting only Find My
without \ac{DCK} 3.0 is still under development.
Apple might add Find My \ac{LPM} support to further devices in the future.

\begin{table}[!t]

\renewcommand{\arraystretch}{1.3}
\centering
\footnotesize

\caption{Wireless chips with actively used LPM support.}
\label{tab:chip_list}
\begin{tabular}{l|lc|llc}

\textbf{Series}	& \textbf{NFC + SE}	& \textbf{LPM} & \textbf{Bluetooth/Wi-Fi}	& \textbf{UWB} & \textbf{LPM} \\
\hline
iPhone X\textsc{r}	& NXP SN100			& $\checkmark$	& BCM4347B1				& --			& --	\\
iPhone X\textsc{s}	& NXP SN100			& $\checkmark$	& BCM4377B2				& --			& --	\\
iPhone 11			& NXP SN200			& $\checkmark$	& BCM4378B1				& r1p0 	& $\checkmark$	\\
iPhone SE 2020		& NXP SN200			& $\checkmark$	& BCM4378B1				& -- 			& --	\\
iPhone 12			& NXP SN210			& $\checkmark$	& BCM4387C2				& r1p1 	& $\checkmark$	\\
iPhone 13			& NXP SN210			& $\checkmark$	& BCM4387C2				& r1p2 	& $\checkmark$	\\

\end{tabular}
\end{table}

%!TEX root = ../lpm.tex

\section{Security Analysis}
\label{sec:threat}

In this section, we analyze security of \ac{LPM} features following a layered approach.
First, we check if \ac{LPM} applications are working as intended to ensure the user's
security and safety, and then analyze
the impact of \ac{LPM} on firmware and hardware security.

\subsection{Applications}

\subsubsection{Adversary Model}

On the application layer, we assume an attacker that did not manipulate the firmware
or software on the iPhone. Instead, they want to compromise or use \ac{LPM} features, e.g.,
disable Find My to steal an iPhone or use Express Cards and Keys to steal money or physical assets.
There is a usability aspect, for example, if the legitimate user does not understand
technological limitations of \ac{LPM} features, theft becomes more likely.

\subsubsection{Find My}
\label{sssec:findmythreat}

When observing Find My, we found that \ac{LPM} does not hold its security promises to users. It stops sending advertisements way earlier than expected
and, in many cases, does not send advertisements at all. In the following, we
explain under which conditions advertisements are sent and which problems occurred
during testing.

\begin{figure}[!b]
\includegraphics[width=0.7\columnwidth]{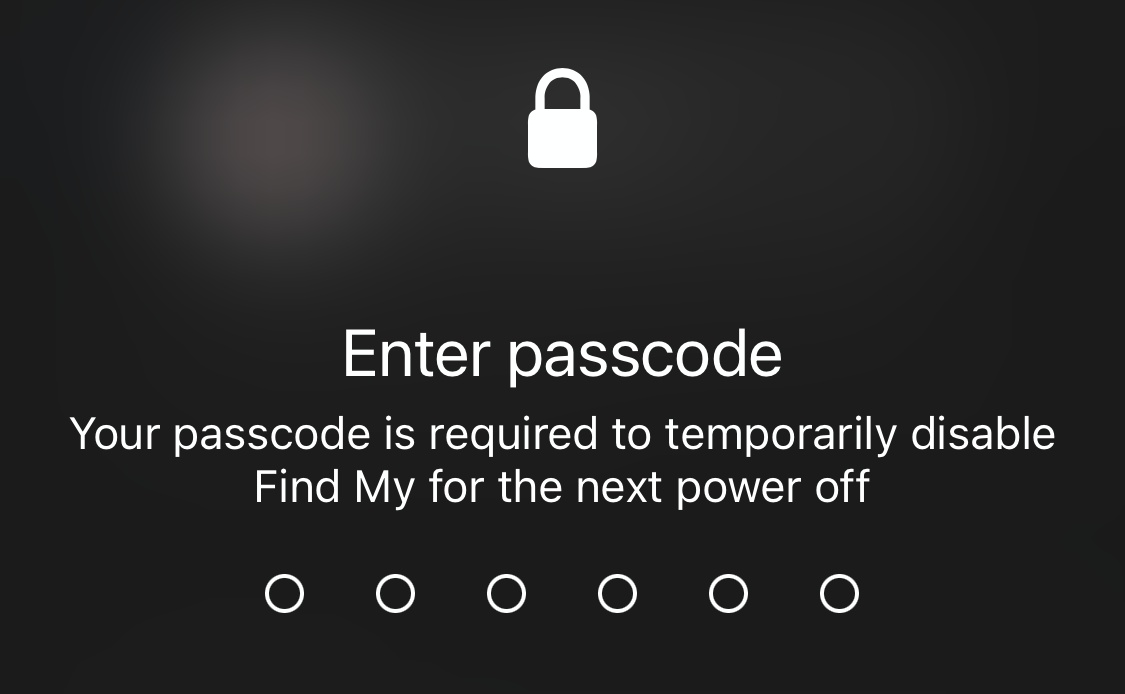}
\caption{Passcode dialogue to disable Find My on power off.}
\label{fig:passcode}
\end{figure}

\paragraph{Maximum \ac{LPM} Time Period}

Only \num{96} advertisements each broadcasted for \SI{15}{\min} are configured
during user-initiated shutdown. After \SI{24}{\hour}, all advertisements are
transmitted. This is not explained to the user, who might think their iPhone
remains findable for multiple days. The current shutdown dialogue suggests
that Find My after power off is a strong security feature. Disabling
even requires a passcode as of iOS 15.3 as additional anti-theft measure,
as shown in \autoref{fig:passcode}.
Users might permanently lose their iPhone when relying too much upon the Find My after power off feature.
The maximum time period could be extended by
configuring more advertisements, cycling advertisements only after \SI{24}{\hour}, or by storing the master beacon key within
an \ac{SE} applet that the Bluetooth chip can query to generate more advertisements.

Upon battery-low shutdown, Find My is only active for slightly more than \SI{5}{\hour}.
This is similar to Express Mode~\cite{expressmode} but not documented at all.
The battery should not be fully discharged to prevent hardware damage. 
Thus, at some point, the iPhone has to switch from power reserve to power off.

\paragraph{Find My Stops Before First Unlock After 24\,h}

On power off, advertisements for the next \SI{24}{\hour} are additionally stored locally on disk.
Even if Find My was disabled on power off, advertisements are still cached.
Once the iPhone boots, it restores a Find My token from an NVRAM storage~\cite[p. 14ff]{iosinternals2},
which survives reboots. The Find My token is then used to decrypt the cached
advertisements, making them accessible to the Bluetooth daemon before first unlock~\cite[p. 98]{appleplatformsec}.
When \SI{24}{\hour} expire, the iPhone stops sending advertisements.
This is irrespective of previous
actions, the iPhone could have been in \ac{LPM} before or rebooted directly.
Since advertisements are fixed to a scheduled transmission time, this period
cannot be extended.

If the iPhone has an Internet connection before first unlock, it connects
to Apple's servers for reporting its location. Many conditions cause the iPhone to be offline before first unlock.
Wi-Fi keys are only available after first unlock~\cite[p. 97]{appleplatformsec},
and a cellular connection fails if the user has set a SIM PIN or the
SIM card was removed.
Thus, a thief can be sure that 
Find My is disabled after \SI{24}{\hour}, even if they power on the iPhone.

\paragraph{Failures Initializing Find My LPM}

Even when the user interface shows that Find My is enabled after power off,
this is sometimes not the case. For example, Find My advertisement generation might fail. We observed this error multiple times
on a device with iOS 15.0.1. If Find My \ac{LPM} setup fails during shutdown, no
warnings are shown to the user.

\paragraph{Unintended Boot}

When connected a power cable is plugged in during user-initiated power off and removed later, the iPhone
boots after a couple of hours. We observed this behavior on two iPhone 12
and an iPhone 13 on iOS 15.3 or lower with Find My enabled.
If no power cable is plugged in during power off, the iPhone does not boot automatically.
On an iPhone 12 on iOS 14, which does not support Find My \ac{LPM}, no unintended boots occurred.
Due to this issue, the user might find their iPhone with an empty battery when Find My
is active, despite switching it off.

\paragraph{Advertisement Accuracy}

We observed a standard deviation of \SI{19}{\second} in the \SI{15}{\minute} rotation windows when capturing advertisements
of an iPhone in \ac{LPM} for \SI{24}{\hour}. This deviation could become an issue
when an iPhone would support \ac{LPM} for more than \SI{24}{\hour}. When the advertisements are outside of
the search window, the Find My servers will no longer find the device.

\todo{add a box plot here and explain windows. are the 19s as maximum outside of the 15min slots even at the end
of the measurement? do they add up?}

\subsubsection{Express Cards and Keys}

The obvious threat for Express Cards and Keys is that a user's credit card or
car key is used without further authentication when their iPhone is stolen. To this end, the user
must actively enable Express Mode once for each item in the Wallet application. Limitations, such as unsupported cards or
shutdown after \SI{5}{\hour} are publicly documented~\cite{expressmode}.
Thus, we assume that most users understand the functionality and risks of Express Mode.

\subsection{Firmware}

\subsubsection{Adversary Model}

On the firmware layer, we assume an attacker with privileged firmware access.
For example, the attacker could be able to send custom commands to the firmware
via the operating system, modify the firmware image, or gain code execution over-the-air.
This is a strong precondition, which we discuss first.
Then, we discuss what an attacker could do once firmware is compromised.

\subsubsection{Tampering with Firmware}

An attacker has different possibilities to gain firmware control, which
also vary in their preconditions.

\paragraph{Local Firmware Modification}

An attacker with system-level access could modify firmware of any component that supports \ac{LPM}.
This way, they maintain (limited) control of the iPhone even when the user powers it off.
This might be interesting for persistent exploits used against high-value targets, such as journalists~\cite{Huang2016Against}.

As of iOS 15, \ac{LPM}-supported chips are \ac{NFC}, \ac{UWB}, and Bluetooth.
An attacker does not need to change kernel behavior and can use existing
drivers to enable \ac{LPM} on these chips, and only needs to modify the chip's firmware.
The \ac{NFC} chip is the only wireless chip on an iPhone that has encrypted and
signed firmware. Even though attempts were made to bypass the secure bootloader, these
were not successful~\cite{nxppentest}. The firmware of the \ac{UWB} chip also has
secure boot but is only signed and not encrypted~\cite{u1-bh,airtag-hwio}. However, the Bluetooth firmware is neither signed nor encrypted.
Since the Bluetooth chip does not have secure boot enabled, it misses
a root of trust that could be used to verify the firmware it loads.
In \autoref{sec:malware}, we show that it is possible to create malware that runs on
iPhone 13 Bluetooth chips, even if the phone is powered off.

\paragraph{Remote Code Execution}

Attackers without system-level access could try to gain code execution on an \ac{LPM}-enabled chip
over the air. In the past, various vulnerabilities have been disclosed for the Bluetooth chip
series used in iPhones~\cite{mantz2019internalblue,frankenstein,braktooth}. Such vulnerabilities could also exist
in other wireless chips.

Minimizing functionality of a wireless stack also reduces its attack surface~\cite{lightblue}.
The Find My module only transmits advertisements over Bluetooth but does not receive data.
However, \ac{NFC} Express Mode as well as Bluetooth and \ac{UWB} \ac{DCK} 3.0 allow
data exchange. Apple already minimizes the attack surface by only enabling these features on demand.

\subsubsection{Reconfiguring Firmware Options}

Even if all firmware would be protected against manipulation, an attacker with system-level access
can still send custom commands to chips.
The Bluetooth daemon configures Find My \ac{LPM} on shutdown with vendor-specific \ac{HCI} commands.
These commands allow a very fine-grained configuration, including advertisement rotation intervals and contents. 
With these commands, an attacker, for example, could set every second advertisement to a public key that matches their account.
Afterward, the attacker could locate the victim device, but the legitimate user would still see their iPhone with a
recent location in Find My.

This specific attack is not relevant as of now, because the Bluetooth firmware on iPhones can be
modified. If Bluetooth chips on the next generation of iPhones would support secure boot,
this attack surface would become relevant. The Apple Watch, which runs an iOS derivative,
has a different Bluetooth chip supporting secure boot~\cite{magicpairing}. If Find My \ac{LPM} would
be added to the Watch, fine-grained configurability would introduce the same issues.

\subsubsection{Direct \acl{SE} Access via Firmware}

Bluetooth and \ac{UWB} \ac{LPM} were introduced to support car keys,
which store their data in an \ac{SE} applet.
Transactions between wireless components and the \ac{SE} should be hidden from
iOS~\cite[p. 141]{appleplatformsec}, ensuring security even if iOS or applications are compromised.
For example, a malicious application that can access the \ac{SE} could relay payment transactions~\cite{serelay}.
Since the Bluetooth firmware can be manipulated, an additional \ac{SE} interface is directly exposed
to iOS. This weakens the trust model for car keys and other
secret information in the \ac{SE}.

Additional interfaces influence the overall security level
provided by the \ac{SE}. It is unclear if the \ac{SE} still holds the same security
standards, because the interface features, applet implementations, and intended
security boundaries are not public.

\subsection{Hardware}
\label{ssec:threathw}

\subsubsection{Adversary Model}

On the hardware layer, we assume that neither attackers nor legitimate users would manipulate the hardware.
We analyze which components could be stealthily powered on while the iPhone is off and which
applications attackers could build. We discuss if users could detect such \ac{LPM} applications.

\subsubsection{Suitable Chips}

Any chip that is controlled by the \acf{PMU} could be active in \ac{LPM}.
This property holds for most chips in an iPhone but still does not make every chip ideal
to be operating in \ac{LPM}.
Since \ac{LPM} should not significantly drain the battery, suitable chips should either
be standalone or combinations of a few battery-friendly chips.
Moreover, these chips should have some way to interact with the user or the outside world.

Even though most chips on an iPhone are connected to the \ac{PMU}, they are also
connected to the \ac{AP} or the \ac{AoP}.
The \ac{AP} is the main processor running iOS. 
The \ac{AoP} is a co-processor running Apple's RTKitOS, an embedded \ac{RTOS}~\cite[p. 20ff]{iosinternals2}.
While the \ac{AoP} is always active and waits for hardware events and triggers, the
\ac{AP} can go to sleep, significantly reducing power consumption.
Drivers in the iOS kernel running on the \ac{AP} often have a companion running on
the \ac{AoP}.
This hardware architecture requires most chips to either interact with the \ac{AoP},
which then routes information to the \ac{AP}, or chips talk directly to the \ac{AP}.
Neither the \ac{AoP} nor the \ac{AP} should be running in \ac{LPM}, because they
consume too much energy.

All wireless chips are connected to the \ac{PMU} and could operate standalone.
This includes the Wi-Fi chip and the cellular baseband, which are currently not used for \ac{LPM} applications.
The display is connected to the \ac{PMU}. In power reserve mode, the display only shows
a static screen. The \ac{AP} sets an image
on the display when entering \ac{LPM}, but wireless chips cannot access the display. Buttons are connected to the \ac{PMU}, such that a button press
can boot or reset the iPhone from any state including \ac{LPM}.
Audio is routed via the \ac{AoP}, meaning that
it is not possible to build a hidden microphone by recording speech and forwarding it to a wireless chip in \ac{LPM}.

\subsubsection{Detection and Deactivation}

Detecting that chips were running an \ac{LPM} firmware
or completely disabling \ac{LPM} is not possible with current iOS on-board tools.
Additional monitoring hardware is needed to detect
wireless transmissions in \ac{LPM}.

\paragraph{Logging}

The original \ac{LPM} firmware images support logging.
The \ac{UWB} chip has a storage for crash logs, including \ac{DCK}, which is read out
after system boot during chip initialization by iOS. The Bluetooth \ac{LPM}
firmware has multiple logging statements, depending on logging flags, which could
be stored on the \ac{SE}. Logs on this level are intended for iOS system developers
and cannot be extracted without jailbreak. Thus, users cannot determine
if \ac{LPM} firmware was running and which actions it performed, even on a system that is not compromised.

\paragraph{Disabling LPM}

\ac{LPM} support is implemented in the hardware. Even if iOS would remove \ac{LPM} features in
the future, these basic hardware features could be activated on a compromised system. 

A possibility to remove permanent support in future hardware revisions would be a physical switch that
disconnects the battery. Apple already has similar physical switches in their hardware,
for example, the microphone on MacBooks is disconnected when the lid is closed~\cite[p. 27]{appleplatformsec}.

\paragraph{Other Protection}

Users who are concerned about being spied on by an iPhone that is off
can install a
transmission monitoring device, which can be mounted on the back
of an iPhone 6 and wiretaps wireless chip connections on the mainboard~\cite{Huang2016Against}.
Such a device would even detect \ac{LPM} malware.
This solution would need to be modified for modern iPhones, if
applicable at all. The mainboard has two stacked and soldered layers since the iPhone 11~\cite{ifixitiphone11}. The desoldering process has a higher risk of damaging the
iPhone than on older models.

Another option would be to put the smartphone into a Faraday bag.
This solution is also considered in \cite{Huang2016Against} but
small holes and gaps can lead to significant leakage.

%!TEX root = ../lpm.tex

\section{Bluetooth Firmware Modification}
\label{sec:malware}

Recent iPhones introduced a new chip interface, including a new Bluetooth firmware patch format.
We show how we can analyze the chip's ROM and RAM nonetheless, and
that the new patch format is unsigned. Based on this, we demonstrate
how to build custom firmware and malware.

\subsection{Dumping and Loading Firmware}

\subsubsection{Firmware Dumps}

We dump the complete firmware for an initial analysis.
Broadcom Bluetooth chips store most of their firmware in ROM.
A full dump including the ROM is required for understanding the firmware.
The ROM is readable with a vendor-specific \acf{HCI} command 
supported by all Broadcom chips, including the one in the iPhone 13.
Apple's Core Bluetooth framework does not allow sending \ac{HCI}
commands to the chip~\cite{corebluetooth}.

On a jailbroken iPhone, we use \emph{InternalBlue} to send
such \ac{HCI} commands~\cite{mantz2019internalblue}. It attaches itself to
the Bluetooth subsystem to send \ac{HCI}
commands and interpret received events.
An iOS-specific behavior of the current InternalBlue version is that the
Bluetooth daemon is detached, and we can exclusively access the chip
without interfering with other system components.
The initial InternalBlue iOS package only supported legacy chips connected
via UART, and we added support for the newer PCIe driver.
With this modification, we dump the ROM of the latest generation
of Bluetooth chips.
For analysis purposes, we dump the RAM in different states.

\subsubsection{Firmware Patching with Patchram}

Firmware is patched temporarily in RAM. 
Broadcom calls this mechanism \emph{Patchram}, and it has been documented within the
InternalBlue project~\cite{mantz2019internalblue}.
%TODO cite https://developer.arm.com/documentation/100166/0001/Debug/Flash-Patch-and-Breakpoint-Unit--FPB-?lang=en ?
%Patchram is based on the ARM \ac{FPB}. 
Any 4-byte value in ROM can be temporarily mapped to another 4-byte value.
On ARM, a function call requires a 4-byte instruction. Thus, a function
in ROM can be overwritten with a function call to an address in RAM.
For this purpose, RAM is writable and executable.
Modern Broadcom chips have 256 Patchram slots~\cite{frankenstein}.
%, and RAM is limited as well.

Legacy chips are connected via UART, and there is no other transport layer than sending
the whole firmware patch using small \ac{HCI} commands with a maximum size of 255\,bytes.
Broadcom's legacy \texttt{.hcd} patch format contains a sequence of
\ac{HCI} commands, which directly write to RAM and configure the Patchram~\cite{msc-dmantz}.

Since the iPhone X\textsc{s}, the Bluetooth chip is connected via PCIe.
This allows for transferring one large \texttt{.bin} patch file at once.
In the following, we reverse-engineer this new format, enabling us to
install custom patches on modern iPhones.

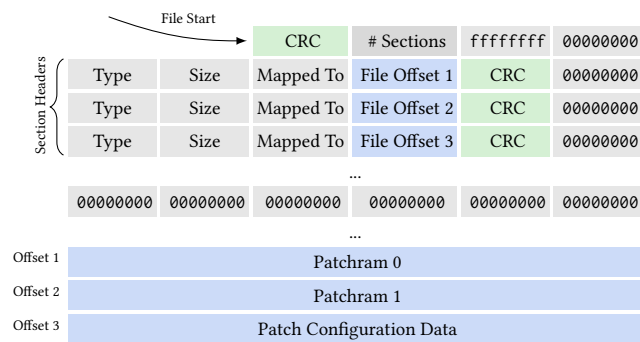
\begin{figure}[!b]

\renewcommand{\arraystretch}{1.3}
% if using array.sty, it might be a good idea to tweak the value of
% \extrarowheight as needed to properly center the text within the cells

\vspace{-1em} %TODO
\centering
%\footnotesize
\setlength\tabcolsep{2pt}
\arrayrulecolor{white}
\setlength{\arrayrulewidth}{1.5pt}

\begin{tikzpicture}[minimum height=0.55cm, scale=0.8, every node/.style={scale=0.8}, node distance=0.7cm]

\node (table) {\begin{tabular}{c|c|c|c|c|c} %TODO adjust overlay

 & & \cellcolor{darkgreen!20} CRC & \cellcolor{gray!30} \# Sections & \cellcolor{gray!20} \texttt{ffffffff} & \cellcolor{gray!20} \texttt{00000000} \\
\hline
\cellcolor{gray!20}Type & \cellcolor{gray!20}Size & \cellcolor{gray!20}Mapped To & \cellcolor{darkblue!20} File Offset 1 & \cellcolor{darkgreen!20} CRC  & \cellcolor{gray!20} \texttt{00000000} \\
\hline
\cellcolor{gray!20}Type & \cellcolor{gray!20}Size & \cellcolor{gray!20}Mapped To & \cellcolor{darkblue!20} File Offset 2 & \cellcolor{darkgreen!20} CRC  & \cellcolor{gray!20} \texttt{00000000} \\
\hline
\cellcolor{gray!20}Type & \cellcolor{gray!20}Size & \cellcolor{gray!20}Mapped To & \cellcolor{darkblue!20} File Offset 3 & \cellcolor{darkgreen!20} CRC  & \cellcolor{gray!20} \texttt{00000000} \\
\multicolumn{6}{c}{...} \\
\cellcolor{gray!20} \texttt{00000000} &\cellcolor{gray!20} \texttt{00000000} &\cellcolor{gray!20} \texttt{00000000} &\cellcolor{gray!20} \texttt{00000000} &\cellcolor{gray!20} \texttt{00000000} &\cellcolor{gray!20} \texttt{00000000} \\
\multicolumn{6}{c}{...} \\
\multicolumn{6}{c}{\cellcolor{darkblue!20} Patchram 0} \\
\hline
\multicolumn{6}{c}{\cellcolor{darkblue!20} Patchram 1} \\
\hline
\multicolumn{6}{c}{\cellcolor{darkblue!20} Patch Configuration Data} \\

\end{tabular}};

% need to shift to the left after adding separators
\begin{scope}[shift={(-0.1, 0)}]

\path[draw] (-4,2.8) edge[bend right=10, ->] node[yshift=0.3cm, xshift=0.2cm] {\footnotesize File Start} (-1.7, 2.4);
\draw [decorate,decoration={brace,amplitude=5pt}] (-4.75,0.5) -- (-4.75,2.1)node [rotate=90,yshift=11,xshift=-22] {\footnotesize Section Headers};

\node[anchor=west, align=left] at (-5.7, -1.2) {\footnotesize{Offset 1}};
\node[anchor=west, align=left] at (-5.7, -1.2-0.56) {\footnotesize{Offset 2}};
\node[anchor=west, align=left] at (-5.7, -1.2-0.56-0.56) {\footnotesize{Offset 3}};

\end{scope}

\end{tikzpicture}

\vspace{-1em} %TODO

\caption{New Bluetooth firmware \texttt{.bin} patch format, introduced with the iPhone X\textsc{s} PCIe Bluetooth chip interface.}
\label{fig:fpi_format}

\end{figure}
%\FloatBarrier

\subsubsection{Firmware Patch Image Upload}

Most communication between iOS and the Bluetooth chip is handled by
the Bluetooth daemon. For selected low-level tasks, the
Bluetooth daemon calls \texttt{BlueTool}. \texttt{BlueTool} allows sending
\ac{HCI} commands to the chip without interfering with the stack, similar to
InternalBlue, and additionally allows loading firmware patches.
\texttt{BlueTool} can be instrumented via \ac{XPC} calls from other daemons, used interactively via a
command line interface, or execute scripts. 
It is an Apple-internal tool without
documentation, and the help output is incomplete. We can load a PCIe firmware as follows:

\begin{lstlisting}
power off
device -D
bcm -w /tmp/path/to/firmware.bin
\end{lstlisting}

Using this method, we can apply firmware patches of arbitrary iOS versions to the chip,
no matter which iOS version a jailbroken research iPhone has.
iOS system logs indicate if a firmware patch image could be loaded or not.

\subsubsection{Firmware Patch Image Format}

RAM dumps enable us to determine address mappings of patches.
We can modify arbitrary parts of the patch file and see if the
chip accepts it. This leads us to the format shown in \autoref{fig:fpi_format}.
The patch is only verified with \acp{CRC} but not signed by Broadcom.

Two Patchram areas in RAM contain writable and executable code regions.
The regular RAM region used by threads to store data is located in between.

A patch configuration data area in \ac{TLV} format defines how patches
are applied. It contains basic information like a human-readable patch name,
hardware configuration settings, as well as 4-byte patch entry instructions
that overwrite the ROM. The format is very similar to the one defined within
\emph{WICED Studio 6.2} \texttt{.hdf} files~\cite{wiced}.
The configuration data section starts with the string \texttt{BRCMcfgS}.
Each patch entry starts with the byte sequence \texttt{0x0110}, is 15\,byte long, and 
contains the ROM address as well as the new 4-byte instruction.
A more detailed description of analyzing and patching configuration data is provided
in our blog post~\cite{unpatchable}.

\subsection{Analyzing and Modifying Firmware}

\subsubsection{Patch Analysis Tooling}

With these new insights, we can load any patch file
on top of an existing InternalBlue ROM dump for static reverse engineering.
We modify existing patch files and load them onto a running
iOS device. We create three scripts to assist us in this process:
\emph{(1)} A standalone Python script, which can extract patch regions
and also reassemble a patch file including correct \acp{CRC}, and \emph{(2)}
an IDA Pro script, which can parse patch configuration data, and \emph{(3)}
another IDA Pro script, which names functions calling the abort handler.

Applications of such tools are very flexible. Broadcom Bluetooth
has been patched in the past to increase \ac{BLE} reliability~\cite{spork2020improving},
enable the chip to send and receive ZigBee~\cite{cayre2021wazabee}, and test against security
issues in the Bluetooth specification~\cite{knob, antonioli20bias}.
In the following, we analyze if the firmware can be modified and loaded into iOS,
and also analyze the internals of the \ac{LPM} module.

%\todo{bluetool and bluetool scripts to load testing firmware}

%\todo{explain resulting .bin format in figure}

\subsubsection{Malware Patch Example}

Preventing firmware tampering during runtime is an important security barrier.
For example, it prevents the Bluetooth daemon from directly observing and modifying communication between
the \ac{SE} and the Bluetooth chip when a \ac{DCK} 3.0 is used. It also hardens against
other inter-chip escalation strategies, e.g., executing code in Wi-Fi via Bluetooth~\cite{spectra}.

iOS can write to the executable RAM regions using a vendor-specific \ac{HCI}
command. This is a legacy feature required for applying \texttt{.hcd} files,
still present in the iPhone 13 Bluetooth chip. The chip protects itself from
modification during runtime with a patch that disables this \ac{HCI} command.
We alter this patch to demonstrate that firmware modification is still possible.

\ac{HCI} commands are handled in the function \path{bt_hci_ingress_HandleCommand}.
This function has an abort handler and is already named by our IDA Pro script.
One of the functions it calls is an \ac{HCI} command filter. The filter has a very specific
pattern and compares \ac{HCI} opcodes. The opcode for writing to RAM is
\texttt{0xfc4c}. Opcodes are compared after subtracting \texttt{0xfc00}.
On an iPhone 12 with a patch level of iOS 15.2, a part of this patch looks as follows:

\begin{lstlisting}[language=ASM]
patch1:002c57d2    cmp  r2, #0x2e ; Download Minidriver
patch1:002c57d4    beq  @\textcolor{gray}{command\_disallowed}@
patch1:002c57d8    cmp  r2, #0x4c ; Write RAM
patch1:002c57da    beq  @\textcolor{gray}{command\_disallowed}@
\end{lstlisting}

After replacing the \texttt{cmp r2, \#0x4c} instruction, repacking the firmware
patch image with the correct \acp{CRC}, and loading it with \texttt{BlueTool},
we can write to the Bluetooth chip RAM during runtime.
As of January 2022, there is no iOS 15.2 jailbreak available. Thus, we load
modified iOS 15.2 firmware to an iPhone SE 2020 with iOS 14.8 and an iPhone
12 with iOS 14.2.1 to confirm our patches work on the chips in the last three generations
of iPhones (see \autoref{tab:chip_list}).

%\todo{patch example: remove writeram protection. now we can tamper with fw during runtime.
%there is no good way to check such tampering in general. the rom can be changed any time,
%also temporarily while a check is performed, etc. secure boot and signed updates are
%state of the art for other wireless chips, such as the nfc and u1 chip in the iphone.}
%\todo{bypasses SE guarantee}

\subsubsection{Custom Bluetooth LPM Applications}

Using the same methods, an attacker could alter the \ac{LPM} application thread to embed malware.
For instance, they could advise the iPhone to send the legitimate owner's Find My advertisements and
additionally send advertisements that make the iPhone locatable for the attacker.
Instead of changing existing functionality, they could also add completely new features.

Broadcom Bluetooth chips can run a standalone application without being connected
to a host. This configuration is very common for IoT devices---a Bluetooth application
like a \ac{BLE} thermal sensor service can run on a single chip.
WICED Studio provides a C wrapper for creating custom standalone applications~\cite{wiced}.
Standalone applications run in a thread called \emph{MPAF}.
Apple's \ac{LPM} application runs in the same thread.
Even though WICED Studio is for Cypress chips, their code base is similar to Broadcom chips~\cite{polypyus},
and we can use it as a base for reverse-engineering and function naming.

The application thread executes functions from a queue.
Functions can be added by calling \texttt{mpaf\_thread\_PostMessage(void *function)}.
For example, Apple's \ac{LPM} application starts by enqueueing a function
that initializes the application thread, and later on, a state machine decides if Find My
advertisements or scanning for cars should be activated.
Functions can also be executed after a timer expires, using 
\path{mpaf_timer_start(timer} \path{*p_tle,} \texttt{int seconds)} with a struct that contains a callback function.
For example, timers are used to cycle Find My advertisements.
Knowing these semantics of the application thread, an attacker
can add custom \ac{LPM} features.

Providing a full analysis of the \ac{LPM} firmware and its modules is out of
scope of the paper. We reverse-engineered a lot of internals, such as
Find My advertisements including their storage and scheduling,
the \ac{GATT} service for \ac{DCK} 3.0 and the \ac{I2C} connection to the \ac{SE}.
The manually reverse-engineered symbols for the iPhone 12 and 13
firmware, including more than checked \num{2000} function names, are published within the \emph{Frankenstein} repository.

%\todo{describe mpaf. only called application thread by cypress.}
%\todo{basic mechanism: offload a function to the thread by calling  }
%\path{mpaf_thread_PostMessage(void *function)}
%\todo{also scheduling sth, this is used for setting up periodic advertisements}
%\path{mpaf_timer_start(tle *p_tle)
%
%\todo{so we can modify existing threads or also add our own stuff}
%
%\todo{the last comand is 0xfe62 0x04 without arguments initializes the mpaf thread. but not sure if relevant here.}
%
%\todo{some of these functions need to be determined by diffing against older firmware. reference polypyus paper.}

%!TEX root = ../lpm.tex

\section{Reverse Engineering Methods}
\label{sec:reversing}

The reverse-engineering techniques we used are also applicable when analyzing other functionality close to hardware on an iPhone.
In the following, we document them, which also ensures reproducibility of our results.

\subsection{Dynamic Analysis}

Dynamic analysis close to hardware requires access to physical devices.
As of now, there is no full iOS emulator, and even commercial solutions
mock up the lower hardware layers~\cite{corellium}.
Ideally, the devices are on different iOS versions and have different
hardware revisions. Using outdated iOS versions is required to use
jailbreaks~\cite{unc0ver,checkra1n}, however, devices might behave differently without 
jailbreak and on the latest iOS release.
The initial efforts to build a proper testing lab meeting these
conditions is nonetheless worth the efforts.

\subsubsection{System Log Analysis}

iOS system logs are very verbose and show how daemons interact with each other.
For example, they tell
which chips go into \ac{LPM} on shutdown, with which \ac{HCI} commands the Find My
\ac{LPM} module is initialized, and that Find My restores beacons from NVRAM before first unlock.

System logs can be extracted from any iOS device, even if not jailbroken.
Apple's debug profiles increase log verbosity~\cite{profileslogs}.
In addition to the Bluetooth profile, we also installed the Location Services profile,
which increases verbosity on some Find My and ranging-related messages.
Logs can be collected up to three days in retrospect, depending on the log type, by running a \emph{sysdiagnose}.
Pressing all buttons until the iPhone vibrates triggers a sysdiagnose, which is
then located a few moments later in \emph{Settings $\rightarrow$ Privacy $\rightarrow$ Analytics Data}.
After unpacking on macOS, the \path{system_logs.logarchive} can be read with the Console app.
This method allows to collect logs including the initial boot process and the very
last messages logged during shutdown.

For our analysis, we created logs under various conditions. For example, we looked
into shutdown logs with Find My enabled and disabled, with and without Express Cards,
user-initiated shutdown versus battery-low shutdown, and combinations of those.
In many cases, these logs are so verbose that it is hard to find relevant messages.
On a device that is actively used, these logs contain multiple millions of messages
just for three days. Messages can be filtered by daemons and log message contents.
The most relevant daemons during our research were \texttt{bluetoothd} for Bluetooth, 
\texttt{nearbyd} for \ac{UWB}, \texttt{nfcd} for \ac{NFC}, \texttt{seserviced} for
the \ac{SE} and \texttt{findmydeviced} for Find My. Naming of features in log messages
is often inconsistent. For example, \ac{LPM} is also called \emph{LPEM} or \emph{LEPM},
and the \ac{DCK} protocol is internally named \emph{Alisha}. This turns log analysis
into a time-consuming and manual process. Also, not everything implemented within
a daemon is logged, and often, conditions under which something interesting is logged
are hard to reach. This is especially the case for the \ac{DCK} protocol, since the 
cars supporting \ac{DCK} 3.0 are yet to be released. Still, there are some hints, such as the \ac{UWB} chip that
attempts to enter \ac{LPM} on power off, and the \texttt{seserviced} checking if
\ac{UWB} Express is active on battery-low shutdown when other \ac{LPM} features are disabled.

\subsubsection{Find My Advertisement Capturing}

Information obtained via log messages should be further analyzed to be confirmed.
For wireless protocols, signals and their timings can be captured and compared to logs
to see if anything is missing. System logs are not available in \ac{LPM},
which is the focus of our research.

One example for missing information in logs is Find My.
Based on log messages, only \num{80} Find My advertisements are set up during power off
via \ac{HCI} commands. As we found later, \num{96} advertisements are configured,
however, \ac{HCI} commands are truncated in the logs.
The logs do not show how often
advertisements are cycled.

We create a Python script running on Linux collecting Find My advertisements.
This script allows us to see when an iPhone starts or stops sending advertisements.
Even if other devices are around, we can assign Bluetooth MAC addresses to devices
by comparing them to system logs. This script assisted us in finding the issues
in \autoref{sssec:findmythreat}. Besides these issues, Find My works very reliable.

\subsubsection{Bluetooth Firmware Instrumentation}

Even though not all iOS versions are jailbreakable, we can backport specific features
for analysis. For example, we can use this to load Bluetooth firmware with \ac{LPM}
support onto a jailbroken device, as explained in \autoref{sec:malware}. Then, we
analyze it dynamically with InternalBlue~\cite{mantz2019internalblue}.

\begin{figure*}[!t]

\renewcommand{\arraystretch}{1.3}
% if using array.sty, it might be a good idea to tweak the value of
% \extrarowheight as needed to properly center the text within the cells

%\vspace{-1em} %TODO
\centering
%\footnotesize
\setlength\tabcolsep{2pt}
\arrayrulecolor{white}
\setlength{\arrayrulewidth}{1.5pt}

\begin{tikzpicture}[minimum height=0.55cm, scale=0.8, every node/.style={scale=0.8}, node distance=0.7cm]

\node (table) {\begin{tabular}{r|l|c|c|l|l|l} %TODO adjust overlay

% & & \cellcolor{darkgreen!20} CRC & \cellcolor{gray!30} \# Sections & \cellcolor{gray!20} \texttt{ffffffff} & \cellcolor{gray!20} \texttt{00000000} \\
%   &  & HCI Opcode \\ %& LPM Command & LPM Parameters \\
1. & Chip Reset \hspace*{3em} & \cellcolor{darkgreen!20} \texttt{0c03} \\ % & LPM Command & LPM Parameters \\
\hline
2. & Find My Config & \cellcolor{darkgreen!20} \texttt{fe62} & \cellcolor{darkblue!20} \texttt{05} & \cellcolor{darkgray!20} \texttt{00800c
0060 00 0f 6000 a005 0000 000003010007} \\
\hline
3. & Advertisements & \cellcolor{darkgreen!20} \texttt{fe62} & \cellcolor{darkblue!20} \texttt{06} & \cellcolor{darkgray!20} \texttt{00000006 [bdbdbdbdbdbd1f1eff4c00121900adadadadadadadadadadadadadadadadadadadadadadad00]}$\times 6$ \\
\hline
 &  & \cellcolor{darkgreen!20} \texttt{fe62} & \cellcolor{darkblue!20} \texttt{06} & \cellcolor{darkgray!20} \texttt{000c0006 [cafecafecafe1f1eff4c00121900ad23ad23ad23ad23ad23ad23ad23ad23ad23ad23ad23ad00]}$\times 6$ \\
& & ... \\
 &  & \cellcolor{darkgreen!20} \texttt{fe62} & \cellcolor{darkblue!20} \texttt{06} & \cellcolor{darkgray!20} \texttt{005a0006 [f00df00df00d1f1eff4c00121900ad42ad42ad42ad42ad42ad42ad42ad42ad42ad42ad42ad00]$\times 6$} \\
\hline
4. & LPM Flags  & \cellcolor{darkgreen!20} \texttt{fe62} & \cellcolor{darkblue!20} \texttt{07} & \cellcolor{darkgray!20} \texttt{0001} \\
\hline
5. & Enter LPM  & \cellcolor{darkgreen!20} \texttt{fe62} & \cellcolor{darkblue!20} \texttt{04} \\
\end{tabular}};

% need to shift to the left after adding separators
%\begin{scope}[shift={(-0.1, 0)}]

\path[draw] (-7.5,2.5) edge[bend left=10, ->] node[yshift=0.3cm, xshift=-0.2cm] {\footnotesize HCI Opcode} (-6.3, 2.2);
\path[draw] (-5.5,2.2) edge[bend left=10, ->] node[yshift=0.45cm, xshift=-0.2cm] {\footnotesize LPM Opcode} (-5.15, 1.65);
\path[draw] (-4.3,2) edge[bend left=10, ->] node[yshift=0.3cm, xshift=-0.1cm] {\footnotesize Total \# of Adv.} (-3.5, 1.65);

%\path[draw] (-1.8-0.5,2) edge[bend right=10, ->] node[yshift=0.27cm, xshift=0cm] {\footnotesize Logging} (-2.8, 1.65);
\path[draw] (-1.2-0.5,2) edge[bend right=10, ->] node[yshift=0.4cm, xshift=0.1cm] {\footnotesize 15\,min} (-2.2, 1.65);
\path[draw] (-0.6,2) edge[bend right=10, ->] node[yshift=0.3cm, xshift=0.3cm] {\footnotesize \# Short Adv.} (-1.6, 1.65);
\path[draw] (0.2,1.9) edge[bend right=10, ->] node[yshift=0.25cm, xshift=0.5cm] {\footnotesize 24\,h} (-0.8, 1.65);
%\path[draw] (1.7,1.8) edge[bend right=10, ->] node[yshift=0.3cm, xshift=0.3cm] {\footnotesize \# Long Adv.} (0.2, 1.65);

\draw [decorate,decoration={brace,amplitude=5pt}] (9.5,1) -- (9.5,-1)node [xshift=38, yshift=29] {\footnotesize 96 Advertisements};
\draw [decorate,decoration={brace,amplitude=3pt}] (-1.25,-0.05) -- (-3.25,-0.05)node [xshift=28, yshift=-9] {\footnotesize MAC};
\draw [decorate,decoration={brace,amplitude=3pt}] (8.5,-0.05) -- (-1,-0.05)node [xshift=135, yshift=-9] {\footnotesize Payload};

% white table overlays because splitting cells has other effects
\fill[fill=white] (2.4,1.05) rectangle ++(7.5,0.8);
\fill[fill=white] (-4.1,-1.85) rectangle ++(14,0.8);
\fill[fill=white] (-3,1.05) rectangle ++(1.5pt,0.6);
\fill[fill=white] (-3.78,1.05) rectangle ++(1.5pt,0.6);
\fill[fill=white] (-2.5,1.05) rectangle ++(1.5pt,0.6);
\fill[fill=white] (-2.05,1.05) rectangle ++(1.5pt,0.6);
\fill[fill=white] (-1.25,1.05) rectangle ++(1.5pt,0.6);
\fill[fill=white] (-0.45,1.05) rectangle ++(1.5pt,0.6);
\fill[fill=white] (0.355,1.05) rectangle ++(1.5pt,0.6);
\fill[fill=white] (-3.45,-1.1) rectangle ++(1.5pt,2.15);

%\end{scope}

\end{tikzpicture}

\vspace{-1em} %TODO

\caption{Configuration of Find My LPM with HCI commands.}
\label{fig:lpm_config}

\end{figure*}

First of all, we can send all \ac{HCI} commands that the Bluetooth daemon sends
when entering \ac{LPM} as well.
These commands are also shown in \autoref{fig:lpm_config}.
When we send the last command, 
it terminates communication with the host. Thus, we cannot execute it during
dynamic analysis of the \ac{LPM} module.
However, we can send all other commands and then take a RAM dump.
This way, we can see in which memory area advertisements and other configuration
is stored and look for functions that reference these.
We can then analyze if some of these values are used as timers or conditional flags.
For example, the \ac{HCI} command \texttt{0xfe62 05 \textcolor{gray}{[19b config]}} contains
a \SI{24}{\hour} timer as minutes in reverse byte order, which is the total time of advertisements,
as well as the total number of advertisements, and the \SI{15}{\min} rotation interval. Thus, we know that these are configurable values.

\subsection{Static Analysis}

Static analysis is often more time-consuming than dynamic analysis. It is still an
important technique, because not everything shows up in logs, and not all functions
are called during runtime. Thus, it can reveal hidden features and shows
significantly more details.

\subsubsection{Hardware Schematics}

There is a large community sharing smartphone schematics, mainly driven by the need
to repair iPhones and other devices independent from vendors. Hence, many schematics
are leaked and publicly available on the Internet.

We use these schematics to confirm that the \ac{SE} is hardwired to \ac{UWB}
and Bluetooth in \autoref{ssec:dck3}. Schematics allow us tracing
connections between other chips to estimate threats by \ac{LPM} in \autoref{ssec:threathw}.
The schematics also contain the interfaces' protocols, e.g., PCIe or \ac{I2C}.
This further helps when reverse-engineering firmware and drivers implementing these protocols.

\subsubsection{System Daemon and Binary Analysis on iOS}

In the following, we first explain how to obtain binaries for iOS, because not all
steps for these are documented comprehensively for recent versions, and then explain
how to analyze them.

\paragraph{Obtaining Binaries}
Analyzing different iOS versions allows to compare features across different software
and hardware revisions.
Even without a physical iPhone, iOS system updates can be downloaded~\cite{ipsw},
unpacked as \texttt{.zip}, and the included \texttt{.dmg} partitions can be mounted.

Recent iOS versions have a \ac{DYLD} shared cache~\cite{dyld}, containing a lot of proprietary
functionality also used by daemons, which is one \SI{4}{\giga\byte}
large binary blob. The packing format changes regularly, and, thus, is often not fully supported by
free reverse engineering tools. After connecting an iPhone to Xcode, an unpacked format containing
separate libraries can be found in the Xcode \texttt{iOS DeviceSupport} folder.

The kernel cache contains the iOS kernel and its drivers. On recent iOS versions, it is no
longer encrypted and can be unpacked with \texttt{img4tool}~\cite{img4tool}.
Sometimes, Apple forgets to strip symbols from the kernel cache.
For example, \ac{OTA} updates~\cite{iosota} for all iPhones and further devices between
iOS 15.0 and iOS 15.1 Beta 3 contain a kernel cache with symbols.
Names in driver interfaces reveal hardware capabilities that are otherwise difficult to
reverse-engineer.

\paragraph{Analysis}

Similar to the approach when analyzing logs, relevant binaries and kernel drivers can be identified by searching
for strings. Log messages seen during dynamic analysis can be searched in binaries.
Interesting binaries can then be loaded into a disassembler or decompiler~\cite{ghidra, binja, ida}.
Often, these tools do not find all references, and logs can be used to reconstruct call graphs.
Even a short analysis can uncover additional features and code paths not reached when capturing logs.

\subsubsection{UWB Firmware}

In addition to Bluetooth firmware, we also analyzed the \ac{UWB} firmware.
It is stored in a file called \texttt{ftab.bin}, which contains firmware for
a high-level \ac{AP} on the \ac{UWB} chip as well as a second firmware for 
a low-level \ac{DSP}~\cite{u1-bh}. The \ac{UWB} firmware is under active
development and features change regularly. New features are sometimes added for testing
and removed later.

Early versions of the iPhone 11 \ac{UWB} firmware had enhanced logging. These versions also contained debug prints for communication with the \ac{SE}.
A lot of testing and logging functionality was removed in later versions.
\emph{Alisha}, the code name for the \ac{DCK} 3.0 protocol, started appearing in the firmware
in iOS 14.3, and \ac{LPM} support for \ac{UWB} was added to \texttt{nearbyd} in the same release.
These examples show that looking into multiple versions to understand new feature integration
is imperative.

%!TEX root = ../lpm.tex

\section{Related Work}
\label{sec:related}

\paragraph{Wireless Firmware on iPhones}
InternalBlue looked into Broadcom's Bluetooth firmware before~\cite{mantz2019internalblue}.
It does not support the new firmware patch format and did not analyze \ac{LPM} features. 
Additionally, Broadcom's Bluetooth firmware was tested with fuzzing~\cite{frankenstein}.
Apple's \ac{UWB} chip was analyzed with regards to basic driver and firmware functionality~\cite{u1-bh, airtag-hwio,airtag-woot},
as well as practical distance shortening attacks~\cite{ghostpeak}. 
The \ac{NFC} chip was not reverse engineered so far. 
A very similar \ac{NFC} chip by NXP has been analyzed and a security issue in the
bootloader was found~\cite{nxppentest}. The researcher also tried attacking the SN100
chip, but the bootloader is different and not affected.
Common Criteria certification reports of NXP's \ac{NFC} chips are publicly available.
The \ac{SE} component in the SN200 \ac{NFC} chip and the full SN100 \ac{NFC} chip were tested, but no exploitable vulnerabilities were found~\cite{sn100test, sn200test}.
Furthermore, the JavaCard Operating System running on the SN200 and SN210 chips was evaluated~\cite{jcop6}.

\paragraph{Wireless Services on iOS}
Various wireless services and their underlying daemons and protocols have been analyzed on iOS.
This includes Find My~\cite{heinrich2021findmy}, AirDrop~\cite{airdrop}, multiple Continuity services~\cite{continuity}, and Magic Pairing~\cite{magicpairing}.
Security of the iOS Bluetooth cellular baseband daemons has been tested with fuzzing~\cite{toothpicker,aristoteles}.

\paragraph{Inter-Chip Attacks}
The direct connection between the \ac{SE} and multiple wireless chips enables potential inter-chip privilege escalation attacks.
A practical attack that enables code execution in Wi-Fi via Bluetooth has been demonstrated to work on recent iPhones~\cite{spectra}.

\paragraph{Transmission Monitoring}
Airplane mode might not disable all wireless
transmissions on a smartphone under surveillance. Thus, Huang and Snowden developed a separate
hardware-based device, which they attached to the mainboard of an iPhone 6 to
monitor wireless chips~\cite{Huang2016Against}. This generation of smartphones
neither considered \ac{LPM} nor supported \ac{UWB}, adding a different attack angle
and requiring upgrades to the existing hardware.

%!TEX root = ../lpm.tex

\section{Conclusion}
\label{sec:conclusion}

The current \ac{LPM} implementation on Apple iPhones is opaque and adds new threats.
Since \ac{LPM} support is based on the iPhone's hardware, it cannot be
removed with system updates. Thus, it has a long-lasting effect on the
overall iOS security model.
To the best of our knowledge, we are the first who looked into undocumented \ac{LPM} features introduced in iOS 15
and uncover various issues.

Design of \ac{LPM} features seems to be mostly driven by functionality, without
considering threats outside of the intended applications.
Find My after power off turns shutdown iPhones into tracking devices by design,
and the implementation within the Bluetooth firmware is not secured against
manipulation. Tracking properties could stealthily be changed by attackers with system-level access.
Furthermore, modern car key support requires \ac{UWB} in \ac{LPM}.
Bluetooth and \ac{UWB} are now hardwired to the \ac{SE},
used to store car keys and other secrets. Given that
Bluetooth firmware can be manipulated, this exposes \ac{SE} interfaces
to iOS. However, the \ac{SE} is specifically meant to protect secrets under
the condition that iOS and applications running on it could be compromised.

Even though \ac{LPM} applications increase security and safety in many use cases,
Apple should add a hardware-based switch to disconnect the battery.
This would improve the situation for privacy-concerned users and surveillance
targets like journalists.

\section*{Availability}
Our tools for modifying and
analyzing recent Broadcom firmware are published within the \emph{InternalBlue} and \emph{Frankenstein} repositories. This includes a standalone firmware
patcher, IDA Pro scripts for interpreting firmware patch files
and naming functions by debug strings, and symbols
for the iPhone 12 and 13 firmware.

\begin{acks}
We thank Matthew Green for initially rising the question how Find My after power off works, Hector Martin for pointing us at iPhone schematics
to confirm what is possible in \ac{LPM}, Fabian Freyer for the discussions about \ac{LPM}, Florian Kosterhon for proofreading this paper and comparing \ac{UWB} firmware versions,
and Apple for reading our paper prior to publication.

This work has been funded by the German Federal Ministry of Education and Research and the Hessen State Ministry for Higher Education, Research and the Arts within their joint support of the National Research Center for Applied Cybersecurity ATHENE.
\end{acks}

\bibliographystyle{ACM-Reference-Format}
\bibliography{bibfile}


\begin{thebibliography}{58}

%%% the Web address.

\ifx \showCODEN    \undefined \def \showCODEN     #1{\unskip}     \fi
\ifx \showDOI      \undefined \def \showDOI       #1{#1}\fi
\ifx \showISBNx    \undefined \def \showISBNx     #1{\unskip}     \fi
\ifx \showISBNxiii \undefined \def \showISBNxiii  #1{\unskip}     \fi
\ifx \showISSN     \undefined \def \showISSN      #1{\unskip}     \fi
\ifx \showLCCN     \undefined \def \showLCCN      #1{\unskip}     \fi
\ifx \shownote     \undefined \def \shownote      #1{#1}          \fi
\ifx \showarticletitle \undefined \def \showarticletitle #1{#1}   \fi
\ifx \showURL      \undefined \def \showURL       {\relax}        \fi
% invisible to TeX
\providecommand\bibfield[2]{#2}
\providecommand\bibinfo[2]{#2}
\providecommand\natexlab[1]{#1}
\providecommand\showeprint[2][]{arXiv:#2}

\bibitem[\protect\citeauthoryear{Antonioli, Tippenhauer, and
  Rasmussen}{Antonioli et~al\mbox{.}}{2020}]%
        {antonioli20bias}
\bibfield{author}{\bibinfo{person}{Daniele Antonioli},
  \bibinfo{person}{Nils~Ole Tippenhauer}, {and} \bibinfo{person}{Kasper
  Rasmussen}.} \bibinfo{year}{2020}\natexlab{}.
\newblock \showarticletitle{{BIAS: Bluetooth Impersonation AttackS}}. In
  \bibinfo{booktitle}{\emph{IEEE Symposium on Security and Privacy}}.
\newblock


\bibitem[\protect\citeauthoryear{Antonioli, Tippenhauer, and
  Rasmussen}{Antonioli et~al\mbox{.}}{2019}]%
        {knob}
\bibfield{author}{\bibinfo{person}{Daniele Antonioli},
  \bibinfo{person}{Nils~Ole Tippenhauer}, {and} \bibinfo{person}{Kasper~B.
  Rasmussen}.} \bibinfo{year}{2019}\natexlab{}.
\newblock \showarticletitle{{The {KNOB} is Broken: Exploiting Low Entropy in
  the Encryption Key Negotiation Of Bluetooth BR/EDR}}.
  \bibinfo{howpublished}{\url{https://www.usenix.org/conference/usenixsecurity19/presentation/antonioli}}.
  In \bibinfo{booktitle}{\emph{28th {USENIX} Security Symposium ({USENIX}
  Security 19)}}. \bibinfo{publisher}{{USENIX} Association},
  \bibinfo{address}{Santa Clara, CA}, \bibinfo{pages}{1047--1061}.
\newblock
\showISBNx{978-1-939133-06-9}


\bibitem[\protect\citeauthoryear{{Apple}}{{Apple}}{2021a}]%
        {appleplatformsec}
\bibfield{author}{\bibinfo{person}{{Apple}}.} \bibinfo{year}{2021}\natexlab{a}.
\newblock \bibinfo{title}{{Apple Platform Security}}.
\newblock
  \bibinfo{howpublished}{\url{https://manuals.info.apple.com/MANUALS/1000/MA1902/en_US/apple-platform-security-guide.pdf}}.
\newblock


\bibitem[\protect\citeauthoryear{{Apple}}{{Apple}}{2021b}]%
        {appleuwbcarwwdc}
\bibfield{author}{\bibinfo{person}{{Apple}}.} \bibinfo{year}{2021}\natexlab{b}.
\newblock \bibinfo{title}{{Explore UWB-based car keys}}.
\newblock
  \bibinfo{howpublished}{\url{https://developer.apple.com/videos/play/wwdc2021/10084/}}.
\newblock


\bibitem[\protect\citeauthoryear{{Apple}}{{Apple}}{2022a}]%
        {corebluetooth}
\bibfield{author}{\bibinfo{person}{{Apple}}.} \bibinfo{year}{2022}\natexlab{a}.
\newblock \bibinfo{title}{{ Core Bluetoot | Apple Developer Documentation}}.
\newblock
  \bibinfo{howpublished}{\url{https://developer.apple.com/documentation/corebluetooth}}.
\newblock


\bibitem[\protect\citeauthoryear{{Apple}}{{Apple}}{2022b}]%
        {profileslogs}
\bibfield{author}{\bibinfo{person}{{Apple}}.} \bibinfo{year}{2022}\natexlab{b}.
\newblock \bibinfo{title}{{Bug Reporting---Profiles and Logs}}.
\newblock
  \bibinfo{howpublished}{\url{https://developer.apple.com/bug-reporting/profiles-and-logs/}}.
\newblock


\bibitem[\protect\citeauthoryear{{Apple}}{{Apple}}{2022c}]%
        {dyld}
\bibfield{author}{\bibinfo{person}{{Apple}}.} \bibinfo{year}{2022}\natexlab{c}.
\newblock \bibinfo{title}{{Overview of Dynamic Libraries}}.
\newblock
  \bibinfo{howpublished}{\url{https://developer.apple.com/library/archive/documentation/DeveloperTools/Conceptual/DynamicLibraries/100-Articles/OverviewOfDynamicLibraries.html}}.
\newblock


\bibitem[\protect\citeauthoryear{{Apple}}{{Apple}}{2022d}]%
        {expressmode}
\bibfield{author}{\bibinfo{person}{{Apple}}.} \bibinfo{year}{2022}\natexlab{d}.
\newblock \bibinfo{title}{{Use Express Mode with cards, passes, and keys in
  Apple Wallet}}.
\newblock
  \bibinfo{howpublished}{\url{https://support.apple.com/en-us/HT212171}}.
\newblock


\bibitem[\protect\citeauthoryear{{Binary Ninja}}{{Binary Ninja}}{2022}]%
        {binja}
\bibfield{author}{\bibinfo{person}{{Binary Ninja}}.}
  \bibinfo{year}{2022}\natexlab{}.
\newblock \bibinfo{title}{{A New Type of Reversing Platform}}.
\newblock \bibinfo{howpublished}{\url{https://binary.ninja}}.
\newblock


\bibitem[\protect\citeauthoryear{{BMW}}{{BMW}}{2022}]%
        {bmwdigitalkey}
\bibfield{author}{\bibinfo{person}{{BMW}}.} \bibinfo{year}{2022}\natexlab{}.
\newblock
  \bibinfo{howpublished}{\url{https://www.bmw.de/de/topics/service-zubehoer/bmw-connecteddrive/digital-key.html}}.
\newblock


\bibitem[\protect\citeauthoryear{{Car Connectivity Consortium}}{{Car
  Connectivity Consortium}}{2020}]%
        {digitalkeywhitepaper}
\bibfield{author}{\bibinfo{person}{{Car Connectivity Consortium}}.}
  \bibinfo{year}{2020}\natexlab{}.
\newblock \bibinfo{title}{{Digital Key -- The Future of Vehicle Access,
  Whitepaper}}.
\newblock
  \bibinfo{howpublished}{\url{https://global-carconnectivity.org/wp-content/uploads/2020/04/CCC_Digital_Key_2.0.pdf}}.
\newblock


\bibitem[\protect\citeauthoryear{{Car Connectivity Consortium}}{{Car
  Connectivity Consortium}}{2021}]%
        {digitalkey30press}
\bibfield{author}{\bibinfo{person}{{Car Connectivity Consortium}}.}
  \bibinfo{year}{2021}\natexlab{}.
\newblock \bibinfo{title}{{Car Connectivity Consortium Publishes Digital Key
  Release 3.0}}.
\newblock
  \bibinfo{howpublished}{\url{https://carconnectivity.org/press-release/car-connectivity-consortium-publishes-digital-key-release-3-0/}}.
\newblock


\bibitem[\protect\citeauthoryear{Cayre, Galtier, Auriol, Nicomette,
  Ka{\^a}niche, and Marconato}{Cayre et~al\mbox{.}}{2021}]%
        {cayre2021wazabee}
\bibfield{author}{\bibinfo{person}{Romain Cayre}, \bibinfo{person}{Florent
  Galtier}, \bibinfo{person}{Guillaume Auriol}, \bibinfo{person}{Vincent
  Nicomette}, \bibinfo{person}{Mohamed Ka{\^a}niche}, {and}
  \bibinfo{person}{G{\'e}raldine Marconato}.} \bibinfo{year}{2021}\natexlab{}.
\newblock \showarticletitle{{WazaBee: attacking Zigbee networks by diverting
  Bluetooth Low Energy chips}}. In \bibinfo{booktitle}{\emph{IEEE/IFIP
  International Conference on Dependable Systems and Networks (DSN)}}.
\newblock


\bibitem[\protect\citeauthoryear{{Christopher Wade}}{{Christopher
  Wade}}{2021}]%
        {nxppentest}
\bibfield{author}{\bibinfo{person}{{Christopher Wade}}.}
  \bibinfo{year}{2021}\natexlab{}.
\newblock
  \bibinfo{howpublished}{\url{https://www.pentestpartners.com/security-blog/breaking-the-nfc-chips-in-tens-of-millions-of-smart-phones-and-a-few-pos-systems/}}.
\newblock


\bibitem[\protect\citeauthoryear{Classen}{Classen}{2021}]%
        {unpatchable}
\bibfield{author}{\bibinfo{person}{Jiska Classen}.}
  \bibinfo{year}{2021}\natexlab{}.
\newblock
  \bibinfo{howpublished}{\url{https://naehrdine.blogspot.com/2021/01/broadcom-bluetooth-unpatching.html}}.
\newblock


\bibitem[\protect\citeauthoryear{Classen, Freyer, and Roth}{Classen
  et~al\mbox{.}}{2021}]%
        {airtag-hwio}
\bibfield{author}{\bibinfo{person}{Jiska Classen}, \bibinfo{person}{Fabian
  Freyer}, {and} \bibinfo{person}{Thomas Roth}.}
  \bibinfo{year}{2021}\natexlab{}.
\newblock \bibinfo{title}{{Over the Air-Tag: shenanigans with the most
  over-engineered keyfinder}}.
\newblock \bibinfo{howpublished}{Presentation at hardwear.io Netherlands 2021,
  \url{https://hardwear.io/netherlands-2021/speakers/jiska-and-fabian-and-stacksmashing.php}}.
\newblock


\bibitem[\protect\citeauthoryear{Classen, Gringoli, Hermann, and
  Hollick}{Classen et~al\mbox{.}}{2022}]%
        {spectra}
\bibfield{author}{\bibinfo{person}{Jiska Classen}, \bibinfo{person}{Francesco
  Gringoli}, \bibinfo{person}{Michael Hermann}, {and} \bibinfo{person}{Matthias
  Hollick}.} \bibinfo{year}{2022}\natexlab{}.
\newblock \showarticletitle{{Attacks on Wireless Coexistence: Exploiting
  Cross-Technology Performance Features for Inter-Chip Privilege Escalation}}.
\newblock  (\bibinfo{year}{2022}).
\newblock


\bibitem[\protect\citeauthoryear{Classen and Heinrich}{Classen and
  Heinrich}{2021}]%
        {u1-bh}
\bibfield{author}{\bibinfo{person}{Jiska Classen} {and}
  \bibinfo{person}{Alexander Heinrich}.} \bibinfo{year}{2021}\natexlab{}.
\newblock \bibinfo{title}{{Wibbly Wobbly, Timey Wimey – What's Really Inside
  Apple's U1 Chip}}.
\newblock \bibinfo{howpublished}{Presentation at Black Hat USA 2021}.
\newblock


\bibitem[\protect\citeauthoryear{{Corellium}}{{Corellium}}{2022}]%
        {corellium}
\bibfield{author}{\bibinfo{person}{{Corellium}}.}
  \bibinfo{year}{2022}\natexlab{}.
\newblock \bibinfo{title}{{Device configuration features}}.
\newblock
  \bibinfo{howpublished}{\url{https://www.corellium.com/platform/features}}.
\newblock


\bibitem[\protect\citeauthoryear{Corporation}{Corporation}{2022}]%
        {wiced}
\bibfield{author}{\bibinfo{person}{Cypress~Semiconductor Corporation}.}
  \bibinfo{year}{2022}\natexlab{}.
\newblock \bibinfo{title}{{WICED Software}}.
\newblock
  \bibinfo{howpublished}{\url{https://www.cypress.com/products/wiced-software}}.
\newblock


\bibitem[\protect\citeauthoryear{Francillon, Danev, and Capkun}{Francillon
  et~al\mbox{.}}{2011}]%
        {pkesshortening}
\bibfield{author}{\bibinfo{person}{Aur{\'{e}}lien Francillon},
  \bibinfo{person}{Boris Danev}, {and} \bibinfo{person}{Srdjan Capkun}.}
  \bibinfo{year}{2011}\natexlab{}.
\newblock \showarticletitle{{Relay Attacks on Passive Keyless Entry and Start
  Systems in Modern Cars}}. In \bibinfo{booktitle}{\emph{Proceedings of the
  Network and Distributed System Security Symposium, {NDSS} 2011, San Diego,
  California, USA, 6th February - 9th February 2011}}. \bibinfo{publisher}{The
  Internet Society}.
\newblock
\urldef\tempurl%
\url{https://www.ndss-symposium.org/ndss2011/relay-attacks-on-passive-keyless-entry-and-start-systems-in-modern-cars}
\showURL{%
\tempurl}


\bibitem[\protect\citeauthoryear{Francis, Hancke, Mayes, and
  Markantonakis}{Francis et~al\mbox{.}}{2011}]%
        {relay2011}
\bibfield{author}{\bibinfo{person}{Lishoy Francis}, \bibinfo{person}{Gerhard~P.
  Hancke}, \bibinfo{person}{Keith Mayes}, {and} \bibinfo{person}{Konstantinos
  Markantonakis}.} \bibinfo{year}{2011}\natexlab{}.
\newblock \showarticletitle{Practical Relay Attack on Contactless Transactions
  by Using NFC Mobile Phones}.
\newblock \bibinfo{journal}{\emph{{IACR} Cryptol. ePrint Arch.}}
  (\bibinfo{year}{2011}), \bibinfo{pages}{618}.
\newblock
\urldef\tempurl%
\url{http://eprint.iacr.org/2011/618}
\showURL{%
\tempurl}


\bibitem[\protect\citeauthoryear{Friebertsh{\"a}user, Kosterhon, Classen, and
  Hollick}{Friebertsh{\"a}user et~al\mbox{.}}{2021}]%
        {polypyus}
\bibfield{author}{\bibinfo{person}{Jan Friebertsh{\"a}user},
  \bibinfo{person}{Florian Kosterhon}, \bibinfo{person}{Jiska Classen}, {and}
  \bibinfo{person}{Matthias Hollick}.} \bibinfo{year}{2021}\natexlab{}.
\newblock \showarticletitle{{Polypyus--The Firmware Historian}}.
\newblock \bibinfo{journal}{\emph{{Workshop on Binary Analysis Research (BAR)
  2021}}} (\bibinfo{date}{Feb} \bibinfo{year}{2021}).
\newblock


\bibitem[\protect\citeauthoryear{Garbelini, Chattopadhyay, Bedi, Sun, and
  Kurniawan}{Garbelini et~al\mbox{.}}{2021}]%
        {braktooth}
\bibfield{author}{\bibinfo{person}{Matheus~E Garbelini},
  \bibinfo{person}{Sudipta Chattopadhyay}, \bibinfo{person}{Vaibhav Bedi},
  \bibinfo{person}{Sumei Sun}, {and} \bibinfo{person}{Ernest Kurniawan}.}
  \bibinfo{year}{2021}\natexlab{}.
\newblock \bibinfo{title}{{\textsc{BrakTooth}: Causing Havoc on Bluetooth Link
  Manager}}.
\newblock
  \bibinfo{howpublished}{\url{https://asset-group.github.io/disclosures/braktooth/}}.
\newblock


\bibitem[\protect\citeauthoryear{Heinrich, Bittner, and Hollick}{Heinrich
  et~al\mbox{.}}{2022}]%
        {airguard}
\bibfield{author}{\bibinfo{person}{Alexander Heinrich}, \bibinfo{person}{Niklas
  Bittner}, {and} \bibinfo{person}{Matthias Hollick}.}
  \bibinfo{year}{2022}\natexlab{}.
\newblock \showarticletitle{{AirGuard -- Protecting Android Users From Stalking
  Attacks By Apple Find My Devices}}. In \bibinfo{booktitle}{\emph{Proceedings
  of the 15th ACM Conference on Security and Privacy in Wireless and Mobile
  Networks}}. \bibinfo{publisher}{{ACM}}.
\newblock


\bibitem[\protect\citeauthoryear{Heinrich, Stute, Kornhuber, and
  Hollick}{Heinrich et~al\mbox{.}}{2021}]%
        {heinrich2021findmy}
\bibfield{author}{\bibinfo{person}{Alexander Heinrich}, \bibinfo{person}{Milan
  Stute}, \bibinfo{person}{Tim Kornhuber}, {and} \bibinfo{person}{Matthias
  Hollick}.} \bibinfo{year}{2021}\natexlab{}.
\newblock \showarticletitle{{Who Can Find My Devices? Security and Privacy of
  Apple’s Crowd-Sourced Bluetooth Location Tracking System}}.
\newblock \bibinfo{journal}{\emph{Proceedings on Privacy Enhancing
  Technologies}}  \bibinfo{volume}{3} (\bibinfo{year}{2021}),
  \bibinfo{pages}{227--245}.
\newblock


\bibitem[\protect\citeauthoryear{Heinze, Classen, and Hollick}{Heinze
  et~al\mbox{.}}{2020a}]%
        {toothpicker}
\bibfield{author}{\bibinfo{person}{Dennis Heinze}, \bibinfo{person}{Jiska
  Classen}, {and} \bibinfo{person}{Matthias Hollick}.}
  \bibinfo{year}{2020}\natexlab{a}.
\newblock \showarticletitle{{ToothPicker: Apple Picking in the iOS Bluetooth
  Stack}}. In \bibinfo{booktitle}{\emph{14th USENIX Workshop on Offensive
  Technologies (WOOT 20)}}. \bibinfo{publisher}{USENIX Association}.
\newblock
\urldef\tempurl%
\url{https://www.usenix.org/conference/woot20/presentation/heinze}
\showURL{%
\tempurl}


\bibitem[\protect\citeauthoryear{Heinze, Classen, and Rohrbach}{Heinze
  et~al\mbox{.}}{2020b}]%
        {magicpairing}
\bibfield{author}{\bibinfo{person}{Dennis Heinze}, \bibinfo{person}{Jiska
  Classen}, {and} \bibinfo{person}{Felix Rohrbach}.}
  \bibinfo{year}{2020}\natexlab{b}.
\newblock \showarticletitle{{MagicPairing: Apple's Take on Securing Bluetooth
  Peripherals}}. In \bibinfo{booktitle}{\emph{Proceedings of the 13th ACM
  Conference on Security and Privacy in Wireless and Mobile Networks}} (Linz,
  Austria) \emph{(\bibinfo{series}{WiSec '20})}.
  \bibinfo{publisher}{Association for Computing Machinery},
  \bibinfo{address}{New York, NY, USA}, \bibinfo{pages}{111–121}.
\newblock
\showISBNx{9781450380065}
\urldef\tempurl%
\url{https://doi.org/10.1145/3395351.3399343}
\showDOI{\tempurl}


\bibitem[\protect\citeauthoryear{{Hex-Rays}}{{Hex-Rays}}{2022}]%
        {ida}
\bibfield{author}{\bibinfo{person}{{Hex-Rays}}.}
  \bibinfo{year}{2022}\natexlab{}.
\newblock \bibinfo{title}{{IDA Pro}}.
\newblock \bibinfo{howpublished}{\url{https://hex-rays.com/ida-pro/}}.
\newblock


\bibitem[\protect\citeauthoryear{Huang and Snowden}{Huang and Snowden}{2016}]%
        {Huang2016Against}
\bibfield{author}{\bibinfo{person}{bunnie Huang} {and} \bibinfo{person}{Edward
  Snowden}.} \bibinfo{year}{2016}\natexlab{}.
\newblock \showarticletitle{{Against the Law: Countering Lawful Abuses of
  Digital Surveillance}}.
\newblock
  \bibinfo{howpublished}{\url{https://www.tjoe.org/pub/direct-radio-introspection}}.
\newblock \bibinfo{journal}{\emph{The Journal of Open Engineering}}
  (\bibinfo{date}{21 7} \bibinfo{year}{2016}).
\newblock
\urldef\tempurl%
\url{https://doi.org/10.21428/12268}
\showDOI{\tempurl}
\newblock
\shownote{https://www.tjoe.org/pub/direct-radio-introspection}.


\bibitem[\protect\citeauthoryear{{Iceman}}{{Iceman}}{2022}]%
        {proxmark}
\bibfield{author}{\bibinfo{person}{{Iceman}}.} \bibinfo{year}{2022}\natexlab{}.
\newblock \bibinfo{title}{{Proxmark3}}.
\newblock
  \bibinfo{howpublished}{\url{https://github.com/RfidResearchGroup/proxmark3}}.
\newblock


\bibitem[\protect\citeauthoryear{{iFixit}}{{iFixit}}{2019}]%
        {ifixitiphone11}
\bibfield{author}{\bibinfo{person}{{iFixit}}.} \bibinfo{year}{2019}\natexlab{}.
\newblock \bibinfo{title}{{iPhone 11 Teardown}}.
\newblock
  \bibinfo{howpublished}{\url{https://www.ifixit.com/Teardown/iPhone+11+Teardown/126192}}.
\newblock


\bibitem[\protect\citeauthoryear{{Just a Penguin}}{{Just a Penguin}}{2022}]%
        {ipsw}
\bibfield{author}{\bibinfo{person}{{Just a Penguin}}.}
  \bibinfo{year}{2022}\natexlab{}.
\newblock \bibinfo{title}{{IPSW Downloads}}.
\newblock \bibinfo{howpublished}{\url{https://ipsw.me}}.
\newblock


\bibitem[\protect\citeauthoryear{{Kim Jong Cracks}}{{Kim Jong Cracks}}{2022}]%
        {checkra1n}
\bibfield{author}{\bibinfo{person}{{Kim Jong Cracks}}.}
  \bibinfo{year}{2022}\natexlab{}.
\newblock \bibinfo{title}{{checkra1n---iPhone 5s -- iPhone X, iOS 12.3 and
  up}}.
\newblock \bibinfo{howpublished}{\url{https://checkra.in/}}.
\newblock


\bibitem[\protect\citeauthoryear{Klee, Roussos, Maass, and Hollick}{Klee
  et~al\mbox{.}}{2020}]%
        {nfcgatepaper}
\bibfield{author}{\bibinfo{person}{Steffen Klee}, \bibinfo{person}{Alexandros
  Roussos}, \bibinfo{person}{Max Maass}, {and} \bibinfo{person}{Matthias
  Hollick}.} \bibinfo{year}{2020}\natexlab{}.
\newblock \showarticletitle{{NFCGate: Opening the Door for NFC Security
  Research with a Smartphone-Based Toolkit}}. In \bibinfo{booktitle}{\emph{14th
  USENIX Workshop on Offensive Technologies (WOOT 20)}}.
  \bibinfo{publisher}{USENIX Association}.
\newblock
\urldef\tempurl%
\url{https://www.usenix.org/conference/woot20/presentation/klee}
\showURL{%
\tempurl}


\bibitem[\protect\citeauthoryear{Kr{\"o}ll, Kleber, Kargl, Hollick, and
  Classen}{Kr{\"o}ll et~al\mbox{.}}{2021}]%
        {aristoteles}
\bibfield{author}{\bibinfo{person}{Tobias Kr{\"o}ll}, \bibinfo{person}{Stephan
  Kleber}, \bibinfo{person}{Frank Kargl}, \bibinfo{person}{Matthias Hollick},
  {and} \bibinfo{person}{Jiska Classen}.} \bibinfo{year}{2021}\natexlab{}.
\newblock \showarticletitle{{ARIstoteles--Dissecting Apple’s Baseband
  Interface}}. In \bibinfo{booktitle}{\emph{European Symposium on Research in
  Computer Security}}. Springer, \bibinfo{pages}{133--151}.
\newblock


\bibitem[\protect\citeauthoryear{Leu, Camurati, Heinrich, Roeschlin, Anliker,
  Hollick, Capkun, and Classen}{Leu et~al\mbox{.}}{2021}]%
        {ghostpeak}
\bibfield{author}{\bibinfo{person}{Patrick Leu}, \bibinfo{person}{Giovanni
  Camurati}, \bibinfo{person}{Alexander Heinrich}, \bibinfo{person}{Marc
  Roeschlin}, \bibinfo{person}{Claudio Anliker}, \bibinfo{person}{Matthias
  Hollick}, \bibinfo{person}{Srdjan Capkun}, {and} \bibinfo{person}{Jiska
  Classen}.} \bibinfo{year}{2021}\natexlab{}.
\newblock \bibinfo{title}{{Ghost Peak: Practical Distance Reduction Attacks
  Against HRP UWB Ranging}}.
\newblock
  \bibinfo{howpublished}{\url{https://securepositioning.com/ghost-peak/}}.
\newblock


\bibitem[\protect\citeauthoryear{Levin}{Levin}{2020}]%
        {iosinternals2}
\bibfield{author}{\bibinfo{person}{Jonathan Levin}.}
  \bibinfo{year}{2020}\natexlab{}.
\newblock \bibinfo{booktitle}{\emph{{*OS Internals, Volume II, Kernel Mode}}}.
\newblock \bibinfo{address}{North Castle, NY}.
\newblock
\showISBNx{978-0-9910555-7-9}


\bibitem[\protect\citeauthoryear{Mantz}{Mantz}{2018}]%
        {msc-dmantz}
\bibfield{author}{\bibinfo{person}{Dennis Mantz}.}
  \bibinfo{year}{2018}\natexlab{}.
\newblock \emph{\bibinfo{title}{{InternalBlue - A Bluetooth Experimentation
  Framework Based on Mobile Device Reverse Engineering}}}.
\newblock Master thesis. \bibinfo{school}{TU Darmstadt}.
\newblock
\newblock
\shownote{Supervised by Matthias Schu\-lz and Jiska Classen}.


\bibitem[\protect\citeauthoryear{Mantz, Classen, Schulz, and Hollick}{Mantz
  et~al\mbox{.}}{2019}]%
        {mantz2019internalblue}
\bibfield{author}{\bibinfo{person}{Dennis Mantz}, \bibinfo{person}{Jiska
  Classen}, \bibinfo{person}{Matthias Schulz}, {and} \bibinfo{person}{Matthias
  Hollick}.} \bibinfo{year}{2019}\natexlab{}.
\newblock \showarticletitle{{InternalBlue - Bluetooth Binary Patching and
  Experimentation Framework}}. In \bibinfo{booktitle}{\emph{The 17th Annual
  International Conference on Mobile Systems, Applications, and Services
  (MobiSys '19)}}.
\newblock
\urldef\tempurl%
\url{https://doi.org/10.1145/3307334.3326089}
\showDOI{\tempurl}


\bibitem[\protect\citeauthoryear{Mayberry, Fenske, Brown, Martin, Fossaceca,
  Rye, Teplov, and Foppe}{Mayberry et~al\mbox{.}}{2021}]%
        {mayberry}
\bibfield{author}{\bibinfo{person}{Travis Mayberry}, \bibinfo{person}{Ellis
  Fenske}, \bibinfo{person}{Dane Brown}, \bibinfo{person}{Jeremy Martin},
  \bibinfo{person}{Christine Fossaceca}, \bibinfo{person}{Erik~C. Rye},
  \bibinfo{person}{Sam Teplov}, {and} \bibinfo{person}{Lucas Foppe}.}
  \bibinfo{year}{2021}\natexlab{}.
\newblock \showarticletitle{{Who Tracks the Trackers? Circumventing Apple's
  Anti-Tracking Alerts in the Find My Network}}. In
  \bibinfo{booktitle}{\emph{Proceedings of the 20th Workshop on Workshop on
  Privacy in the Electronic Society}} (Virtual Event, Republic of Korea)
  \emph{(\bibinfo{series}{WPES '21})}. \bibinfo{publisher}{Association for
  Computing Machinery}, \bibinfo{address}{New York, NY, USA},
  \bibinfo{pages}{181–186}.
\newblock
\showISBNx{9781450385275}
\urldef\tempurl%
\url{https://doi.org/10.1145/3463676.3485616}
\showDOI{\tempurl}


\bibitem[\protect\citeauthoryear{{National Security Agency}}{{National Security
  Agency}}{2022}]%
        {ghidra}
\bibfield{author}{\bibinfo{person}{{National Security Agency}}.}
  \bibinfo{year}{2022}\natexlab{}.
\newblock \bibinfo{title}{{Ghidra}}.
\newblock \bibinfo{howpublished}{\url{https://ghidra-sre.org/}}.
\newblock


\bibitem[\protect\citeauthoryear{{NXP Semiconductors}}{{NXP
  Semiconductors}}{2020}]%
        {jcop6}
\bibfield{author}{\bibinfo{person}{{NXP Semiconductors}}.}
  \bibinfo{year}{2020}\natexlab{}.
\newblock \bibinfo{title}{{NXP JCOP6.x on SN200.C04, Secure Element, Product
  Evaluation Document}}.
\newblock
  \bibinfo{howpublished}{\url{https://www.commoncriteriaportal.org/files/epfiles/nscib-cc-235773_2-st-lite.pdf}}.
\newblock


\bibitem[\protect\citeauthoryear{Roland, Langer, and Scharinger}{Roland
  et~al\mbox{.}}{2012}]%
        {serelay}
\bibfield{author}{\bibinfo{person}{Michael Roland}, \bibinfo{person}{Josef
  Langer}, {and} \bibinfo{person}{Josef Scharinger}.}
  \bibinfo{year}{2012}\natexlab{}.
\newblock \showarticletitle{Relay Attacks on Secure Element-Enabled Mobile
  Devices}. In \bibinfo{booktitle}{\emph{Information Security and Privacy
  Research}}, \bibfield{editor}{\bibinfo{person}{Dimitris Gritzalis},
  \bibinfo{person}{Steven Furnell}, {and} \bibinfo{person}{Marianthi
  Theoharidou}} (Eds.). \bibinfo{publisher}{Springer Berlin Heidelberg},
  \bibinfo{address}{Berlin, Heidelberg}, \bibinfo{pages}{1--12}.
\newblock
\showISBNx{978-3-642-30436-1}


\bibitem[\protect\citeauthoryear{Roth, Freyer, Hollick, and Classen}{Roth
  et~al\mbox{.}}{2022}]%
        {airtag-woot}
\bibfield{author}{\bibinfo{person}{Thomas Roth}, \bibinfo{person}{Fabian
  Freyer}, \bibinfo{person}{Matthias Hollick}, {and} \bibinfo{person}{Jiska
  Classen}.} \bibinfo{year}{2022}\natexlab{}.
\newblock \showarticletitle{{AirTag of the Clones: Shenanigans with Liberated
  Item Finders}}.
\newblock  (\bibinfo{date}{Aug.} \bibinfo{year}{2022}).
\newblock


\bibitem[\protect\citeauthoryear{Ruge, Classen, Gringoli, and Hollick}{Ruge
  et~al\mbox{.}}{2020}]%
        {frankenstein}
\bibfield{author}{\bibinfo{person}{Jan Ruge}, \bibinfo{person}{Jiska Classen},
  \bibinfo{person}{Francesco Gringoli}, {and} \bibinfo{person}{Matthias
  Hollick}.} \bibinfo{year}{2020}\natexlab{}.
\newblock \showarticletitle{{Frankenstein: Advanced Wireless Fuzzing to Exploit
  New Bluetooth Escalation Targets}}. In \bibinfo{booktitle}{\emph{29th
  {USENIX} Security Symposium ({USENIX} Security 20)}}.
  \bibinfo{publisher}{{USENIX} Association}, \bibinfo{pages}{19--36}.
\newblock
\showISBNx{978-1-939133-17-5}
\urldef\tempurl%
\url{https://www.usenix.org/conference/usenixsecurity20/presentation/ruge}
\showURL{%
\tempurl}


\bibitem[\protect\citeauthoryear{Singh, Roeschlin, Zalzala, Leu, and
  {\v{C}}apkun}{Singh et~al\mbox{.}}{2021}]%
        {wisecuwb}
\bibfield{author}{\bibinfo{person}{Mridula Singh}, \bibinfo{person}{Marc
  Roeschlin}, \bibinfo{person}{Ezzat Zalzala}, \bibinfo{person}{Patrick Leu},
  {and} \bibinfo{person}{Srdjan {\v{C}}apkun}.}
  \bibinfo{year}{2021}\natexlab{}.
\newblock \showarticletitle{{Security Analysis of IEEE 802.15. 4z/HRP UWB
  Time-of-Flight Distance Measurement}}. In
  \bibinfo{booktitle}{\emph{Proceedings of the 14th ACM Conference on Security
  and Privacy in Wireless and Mobile Networks}}. \bibinfo{pages}{227--237}.
\newblock


\bibitem[\protect\citeauthoryear{Sp{\"o}rk, Classen, Boano, Hollick, and
  R{\"o}mer}{Sp{\"o}rk et~al\mbox{.}}{2020}]%
        {spork2020improving}
\bibfield{author}{\bibinfo{person}{Michael Sp{\"o}rk}, \bibinfo{person}{Jiska
  Classen}, \bibinfo{person}{Carlo~Alberto Boano}, \bibinfo{person}{Matthias
  Hollick}, {and} \bibinfo{person}{Kay R{\"o}mer}.}
  \bibinfo{year}{2020}\natexlab{}.
\newblock \showarticletitle{{Improving the Reliability of Bluetooth Low Energy
  Connections.}}. In \bibinfo{booktitle}{\emph{EWSN}}.
  \bibinfo{pages}{144--155}.
\newblock


\bibitem[\protect\citeauthoryear{Staeblein}{Staeblein}{2021}]%
        {nxpdigitalkey}
\bibfield{author}{\bibinfo{person}{Markus Staeblein}.}
  \bibinfo{year}{2021}\natexlab{}.
\newblock \bibinfo{title}{{3 Reasons Why CCC Digital Key Release 3.0 Is Cause
  for Excitement}}.
\newblock
  \bibinfo{howpublished}{\url{https://www.nxp.com/company/blog/3-reasons-why-ccc-digital-key-release-3-0-is-cause-for-excitement:BL-CCC-DIGITAL-KEY}}.
\newblock


\bibitem[\protect\citeauthoryear{Stute, Heinrich, Lorenz, and Hollick}{Stute
  et~al\mbox{.}}{2021}]%
        {continuity}
\bibfield{author}{\bibinfo{person}{Milan Stute}, \bibinfo{person}{Alexander
  Heinrich}, \bibinfo{person}{Jannik Lorenz}, {and} \bibinfo{person}{Matthias
  Hollick}.} \bibinfo{year}{2021}\natexlab{}.
\newblock \showarticletitle{{Disrupting Continuity of Apple{\textquoteright}s
  Wireless Ecosystem Security: New Tracking, DoS, and MitM Attacks on iOS and
  macOS Through Bluetooth Low Energy, AWDL, and Wi-Fi}}. In
  \bibinfo{booktitle}{\emph{30th USENIX Security Symposium (USENIX Security
  21)}}. \bibinfo{publisher}{USENIX Association}, \bibinfo{pages}{3917--3934}.
\newblock
\showISBNx{978-1-939133-24-3}
\urldef\tempurl%
\url{https://www.usenix.org/conference/usenixsecurity21/presentation/stute}
\showURL{%
\tempurl}


\bibitem[\protect\citeauthoryear{Stute, Narain, Mariotto, Heinrich,
  Kreitschmann, Noubir, and Hollick}{Stute et~al\mbox{.}}{2019}]%
        {airdrop}
\bibfield{author}{\bibinfo{person}{Milan Stute}, \bibinfo{person}{Sashank
  Narain}, \bibinfo{person}{Alex Mariotto}, \bibinfo{person}{Alexander
  Heinrich}, \bibinfo{person}{David Kreitschmann}, \bibinfo{person}{Guevara
  Noubir}, {and} \bibinfo{person}{Matthias Hollick}.}
  \bibinfo{year}{2019}\natexlab{}.
\newblock \showarticletitle{{A Billion Open Interfaces for Eve and Mallory:
  MitM, DoS, and Tracking Attacks on iOS and macOS Through Apple Wireless
  Direct Link}}. In \bibinfo{booktitle}{\emph{28th {USENIX} Security Symposium
  ({USENIX} Security 19)}}. \bibinfo{publisher}{{USENIX} Association},
  \bibinfo{address}{Santa Clara, CA}, \bibinfo{pages}{37--54}.
\newblock
\showISBNx{978-1-939133-06-9}
\urldef\tempurl%
\url{https://www.usenix.org/conference/usenixsecurity19/presentation/stute}
\showURL{%
\tempurl}


\bibitem[\protect\citeauthoryear{{The iPhone Wiki}}{{The iPhone Wiki}}{2022}]%
        {iosota}
\bibfield{author}{\bibinfo{person}{{The iPhone Wiki}}.}
  \bibinfo{year}{2022}\natexlab{}.
\newblock \bibinfo{title}{{OTA Updates}}.
\newblock
  \bibinfo{howpublished}{\url{https://www.theiphonewiki.com/wiki/OTA_Updates}}.
\newblock


\bibitem[\protect\citeauthoryear{{tihmstar}}{{tihmstar}}{2022}]%
        {img4tool}
\bibfield{author}{\bibinfo{person}{{tihmstar}}.}
  \bibinfo{year}{2022}\natexlab{}.
\newblock \bibinfo{title}{\texttt{img4tool}}.
\newblock \bibinfo{howpublished}{\url{https://github.com/tihmstar/img4tool}}.
\newblock


\bibitem[\protect\citeauthoryear{{TÜV Rheinland Nederland B.V.}}{{TÜV
  Rheinland Nederland B.V.}}{2019}]%
        {sn200test}
\bibfield{author}{\bibinfo{person}{{TÜV Rheinland Nederland B.V.}}}
  \bibinfo{year}{2019}\natexlab{}.
\newblock \bibinfo{title}{{Certification Report, SN200 Series - Secure Element
  with Crypto Library SN200\_SE B1.1 C04}}.
\newblock
  \bibinfo{howpublished}{\url{https://commoncriteriaportal.org/files/epfiles/NSCIB-CC-217812-CR.pdf}}.
\newblock


\bibitem[\protect\citeauthoryear{{TÜV Rheinland Nederland B.V.}}{{TÜV
  Rheinland Nederland B.V.}}{2021}]%
        {sn100test}
\bibfield{author}{\bibinfo{person}{{TÜV Rheinland Nederland B.V.}}}
  \bibinfo{year}{2021}\natexlab{}.
\newblock \bibinfo{title}{{Certification Report, NXP JCOP 5.2 on SN100.C58
  Secure Element}}.
\newblock
  \bibinfo{howpublished}{\url{https://www.commoncriteriaportal.org/files/epfiles/NSCIB-CC-0023577-CR3.pdf}}.
\newblock


\bibitem[\protect\citeauthoryear{unc0ver}{unc0ver}{2022}]%
        {unc0ver}
\bibfield{author}{\bibinfo{person}{unc0ver}.} \bibinfo{year}{2022}\natexlab{}.
\newblock \bibinfo{title}{{The moast advanced jailbreak tool}}.
\newblock \bibinfo{howpublished}{\url{https://unc0ver.dev}}.
\newblock


\bibitem[\protect\citeauthoryear{{Wikileaks}}{{Wikileaks}}{2014}]%
        {wikileaks}
\bibfield{author}{\bibinfo{person}{{Wikileaks}}.}
  \bibinfo{year}{2014}\natexlab{}.
\newblock \bibinfo{title}{{Weeping Angel (Extending) Engineering Notes}}.
\newblock
  \bibinfo{howpublished}{\url{https://wikileaks.org/ciav7p1/cms/page_12353643.html}}.
\newblock


\bibitem[\protect\citeauthoryear{Wu, Wu, Antonioli, Payer, Tippenhauer, Xu,
  Tian, and Bianchi}{Wu et~al\mbox{.}}{2021}]%
        {lightblue}
\bibfield{author}{\bibinfo{person}{Jianliang Wu}, \bibinfo{person}{Ruoyu Wu},
  \bibinfo{person}{Daniele Antonioli}, \bibinfo{person}{Mathias Payer},
  \bibinfo{person}{Nils~Ole Tippenhauer}, \bibinfo{person}{Dongyan Xu},
  \bibinfo{person}{Dave~(Jing) Tian}, {and} \bibinfo{person}{Antonio Bianchi}.}
  \bibinfo{year}{2021}\natexlab{}.
\newblock \showarticletitle{{LIGHTBLUE: Automatic {Profile-Aware} Debloating of
  Bluetooth Stacks}}. In \bibinfo{booktitle}{\emph{30th USENIX Security
  Symposium (USENIX Security 21)}}. \bibinfo{publisher}{USENIX Association},
  \bibinfo{pages}{339--356}.
\newblock
\showISBNx{978-1-939133-24-3}
\urldef\tempurl%
\url{https://www.usenix.org/conference/usenixsecurity21/presentation/wu-jianliang}
\showURL{%
\tempurl}


\end{thebibliography}

\end{document}